\documentclass[english,twocolumn, prl, twoside, superscriptaddress,nobibnotes]{revtex4-1}
\usepackage[T1]{fontenc}
\usepackage[utf8]{inputenc}
\usepackage{refstyle}
\usepackage{amsmath}
\usepackage{amssymb}
\usepackage{graphicx}
\usepackage{multibib}
\usepackage[pdftex, hidelinks]{hyperref}
\usepackage[dvipsnames]{xcolor}
\usepackage{float}

\newcommand\myshade{85}
\colorlet{mylinkcolor}{NavyBlue}
\colorlet{mycitecolor}{NavyBlue}
\colorlet{myurlcolor}{NavyBlue}
\hypersetup{
	linkcolor  = mylinkcolor!\myshade!black,
	citecolor  = mycitecolor!\myshade!black,
	urlcolor   = myurlcolor!\myshade!black,
	colorlinks = true,
}

\makeatletter
\usepackage{nicefrac}
\usepackage[T1]{fontenc}
\newcommand{\sm}{Supplement 1 }

\makeatother

\usepackage{babel}
\begin{document}
\title{Temporal imaging for ultra-narrowband few-photon states of light} 


\author{Mateusz Mazelanik}
\email{m.mazelanik@cent.uw.edu.pl}
\affiliation{Centre for Quantum Optical Technologies, Centre of New Technologies,
University of Warsaw, Banacha 2c, 02-097 Warsaw, Poland}
\affiliation{Faculty of Physics, University of Warsaw, Pasteura 5, 02-093 Warsaw,
Poland}
\author{Adam Leszczyński}
\affiliation{Centre for Quantum Optical Technologies, Centre of New Technologies,
University of Warsaw, Banacha 2c, 02-097 Warsaw, Poland}
\affiliation{Faculty of Physics, University of Warsaw, Pasteura 5, 02-093 Warsaw,
Poland}
\author{Michał Lipka}
\affiliation{Centre for Quantum Optical Technologies, Centre of New Technologies,
University of Warsaw, Banacha 2c, 02-097 Warsaw, Poland}
\affiliation{Faculty of Physics, University of Warsaw, Pasteura 5, 02-093 Warsaw,
Poland}
\author{Michał Parniak}
\affiliation{Centre for Quantum Optical Technologies, Centre of New Technologies,
University of Warsaw, Banacha 2c, 02-097 Warsaw, Poland}
\affiliation{Niels Bohr Institute, University of Copenhagen, Blegdamsvej 17, DK-2100
Copenhagen, Denmark}
\author{Wojciech Wasilewski}
\affiliation{Centre for Quantum Optical Technologies, Centre of New Technologies,
University of Warsaw, Banacha 2c, 02-097 Warsaw, Poland}

\begin{abstract} Plenty of quantum information protocols are enabled by manipulation and detection of photonic spectro-temporal degrees of freedom via light-matter interfaces. While present implementations are  well suited for high-bandwidth  photon sources such as quantum dots, they lack the high resolution required for intrinsically narrow-band light-atom interactions. Here, we demonstrate far-field temporal imaging based on ac-Stark spatial spin-wave phase manipulation in a multimode gradient echo memory. We achieve spectral resolution of 20 kHz with MHz-level bandwidth and ultra-low noise equivalent to 0.023 photons, enabling operation in the single-quantum regime. \end{abstract}

\maketitle

\section{Introduction}
\begin{figure}[!t]
\includegraphics[width=1\columnwidth]{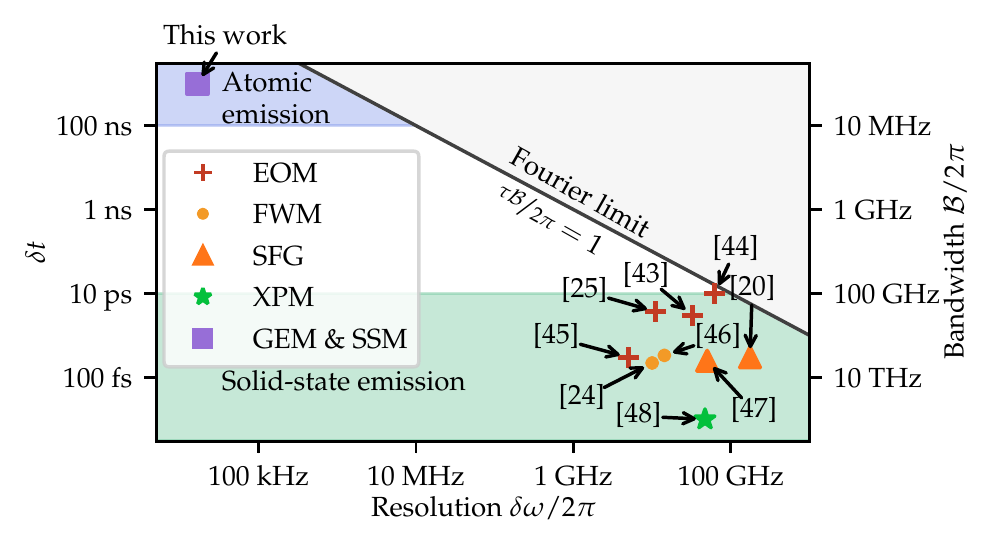}\caption{Temporal imaging state-of-the-art, characterized by temporal $\delta t$
and spectral $\delta\omega$ resolutions. Numerous implementations
based on solid media (Electro-Optic Modulators (EOMS) - \cite{foster_ultrafast_2009,Kauffman1994,Azana2004,Babashah2019},
Four Wave Mixing (FWM) - \cite{foster_silicon-chip-based_2008,Salem2009}, Sum-frequency Generation (SFG) - \cite{bennett_upconversion_1999,Suret2016},
Cross Phase Modulation (XPM) - \cite{Mouradian2000}) are well suited for high bandwidth pico-
or even femtosecond pulses, achieving spectral resolution no better
than $1$ GHz, with time-bandwidth products ($\tau\mathcal{B}$) reaching
$2\pi\times2000$. Our system (Gradient Echo Memory and Spatial Spin-wave Modulation (GEM \& SSM)) has $10^{6}$ times better spectral resolution
$\delta\omega/2\pi\sim$20 kHz, maintaining good $\tau\mathcal{B}$,
thus allowing exploration of previously unattainable region. The grayed region indicates unphysical area bounded by the Fourier limit $\tau\mathcal{B}/2\pi=1$.}
\label{fig:state_of_art}
\end{figure}

The temporal degree of freedom of both classical and quantum states
of light enables or enhances a plethora of quantum information processing
tasks \cite{Humphreys2014,Brecht2015,Reimer2016,Lu2019}. In
the development of quantum network architectures and novel quantum
computing and metrology solutions, a significant effort has been devoted
to quantum memories based on atomic ensembles, offering multi-mode
storage and processing \cite{Pu2017,Parniak2017,mazelanik_coherent_2019,Seri2019},
high efficiency \cite{Cho2016} or long storage-times \cite{Bao2012}.
Feasible implementations of protocols merging the flexibility of atomic
systems and temporal processing capabilities inherently require an
ability to manipulate and detect temporal photonic modes with spectral
and temporal resolution matched to the narrow-band atomic emission.
A versatile approach leveraging spectro-temporal duality, is to perform
a frequency to time mapping -- Fourier transform -- in an analogy
with far-field imaging in position-momentum space. To preserve quantum
structure of non-classical states of light, systems relying on the
concept of a time lens are employed \cite{kolner_temporal_1989,zhu_aberration-corrected_2013,patera_space-time_2018};
however, presently existing physical implementations are well suited
for high-bandwidth systems and involve either electro-optical phase
modulation \cite{kolner_active_1988,grischkowsky_optical_1974,karpinski_bandwidth_2017},
sum-frequency generation \cite{hernandez_104_2013,bennett_aberrations_2001,bennett_temporal_1994,bennett_upconversion_1999,agrawal_temporal_1989}
or four-wave mixing \cite{kuzucu_spectral_2009,okawachi_high-resolution_2009,foster_silicon-chip-based_2008,foster_ultrafast_2009}
in solid-state media. Figure \ref{fig:state_of_art} localizes the existing
schemes in the bandwidth-resolution space. Methods relying on the
time-lensing concept enable spectral shaping \cite{li_high-contrast_2015,donohue_spectrally_2016,Lu2018},
temporal ghost imaging \cite{denis_temporal_2017,dong_long-distance_2016,ryczkowski_magnified_2017,wu_temporal_2019}
and bandwidth matching \cite{allgaier_highly_2017} for photons generated
in dissimilar nodes of a quantum network. While those solutions offer
spectral resolution suitable for high-bandwidth photons generated
in spontaneous parametric down conversion (SPDC) or quantum-dot single-photon
sources, their performance is severely limited in the case of spectrally ultra-narrow
atomic emission ranging from few MHz to tens of kHz \cite{Zhao2014,Guo2017,Farrera2016}, cavity coupled ions (below 100 kHz) \cite{Stute2012}, cavity-enhanced SPDC designed for atomic quantum memories (below 1 MHz) \cite{Rambach2016}, or optomechanical systems \cite{Hong2017, Hill2012}. 
%

In this Article we propose and experimentally demonstrate a novel,
high-spectral-resolution approach to far-field temporal imaging which
is inherently bandwidth-compatible with atomic systems, a regime previously
unexplored as seen in Fig.~\ref{fig:state_of_art}, and works at the
single-photon-level. This allows preservation of quantum correlations, manipulation of field-orthogonal temporal modes \cite{Brecht2015} and characterization of the time-frequency entanglement \cite{Mei2019} of photons from
atomic emission. Our novel technique utilizes a recently developed spin-wave modulation method combined with unusual interpretation of Gradient Echo Memory (GEM) \cite{Hosseini2009} protocol to realize complete temporal imaging setup in a one physical system. 

\section{Principles of temporal imaging}
Imaging systems generally consist of lenses interleaved with free-space
propagation. Analogously, temporal imaging requires an equivalent
of these two basic components. Involved transformations can be viewed
in temporal or spectral domain separately, or equivalently by employing
a spectro-temporal (chronocyclic) Wigner function defined as $W(t,\omega)=1/\sqrt{2\pi}\int_{-\infty}^{\infty}\mathrm{d}\xi A(t+\xi/2)A^{*}(t-\xi/2)\exp(-i\omega\xi)$,
where $A(t)$ denotes the slowly varying amplitude of the signal pulse.

Temporal far-field imaging is typically achieved with a single time-lens preceded
and followed by a temporal analog of free-space propagation.
However, such a setup is equivalent to two lenses interleaved with a single
propagation. In the Wigner function representation, combination of two
temporal lenses with focal lengths $f_{\mathrm{t}}$, separated by
a temporal propagation by the time $f_{\mathrm{t}}$,
is described using a spectro-temporal equivalent of ray transfer matrix:
\begin{equation}
\negthickspace\negthickspace\begin{bmatrix}t'\!\\
\frac{\omega'}{\omega_{0}}\!
\end{bmatrix}\!\negthickspace=\!\negthickspace\begin{bmatrix}1 & 0\\
-\frac{1}{f_{\mathrm{t}}} & 1
\end{bmatrix}\negthickspace\negthickspace\begin{bmatrix}1 & f_{\mathrm{t}}\\
0 & 1
\end{bmatrix}\negthickspace\negthickspace\begin{bmatrix}1 & 0\\
-\frac{1}{f_{\mathrm{t}}} & 1
\end{bmatrix}\negthickspace\negthickspace\begin{bmatrix}t\!\\
\frac{\omega}{\omega_{0}}\!
\end{bmatrix}\!\negthickspace=\!\negthickspace\begin{bmatrix}0 & f_{\mathrm{t}}\\
-\frac{1}{f_{\mathrm{t}}} & 0
\end{bmatrix}\negthickspace\negthickspace\begin{bmatrix}t\!\\
\frac{\omega}{\omega_{0}}\!
\end{bmatrix},\label{eq:phase_space_rot}
\end{equation}
which represents a $\pi/2$ rotation in the phase space, exchanging
temporal and spectral domains, where $\omega_0$ is the optical carrier frequency. To visualize this, in Fig. \ref{fig:setup} (d) we present an equivalent of a cat state Wigner function (corresponding to sequence presented in Fig \ref{fig:sin_fsin} (c)) undergoing these three subsequent transformations: (1) - time-lens, (2) - propagation, (3) - time lens. 

To realize the time-lens with a focal length $f_{\mathrm{t}}$ one has to impose a quadratic phase $\exp[i\omega_{0}t^{2}/(2f_{\mathrm{t}})]$ on the optical pulse $A(t)\to A(t)\exp[i\omega_{0}t^{2}/(2f_{\mathrm{t}})]$, where $\omega_0$ is the optical carier frequency. In the
language of Wigner functions, this transformation can be written as
$W(t,\omega)\to W(t,\omega')$ with $\omega'=\omega-\omega_{0}t/f_{\mathrm{t}}$.
This corresponds to adding to the pulse a linear chirp $\omega(t)=\omega_{0}+\alpha t$.  Typically, such a transformation is achieved by directly modulating the signal pulse using electro-optic modulators \cite{foster_ultrafast_2009,Kauffman1994,Azana2004,Babashah2019}. 

The analog of free-space propagation can be understood as a frequency-dependent delay applied to an optical pulse. In the language of the Wigner function, the transformation takes a form $W(t,\omega)\to W(t',\omega)$ with $t'=t+f_{\mathrm{t}}\omega/\omega_{0}$. Equivalently, a pulse with spectral amplitude $\widetilde{A}(\omega)=\mathcal{F}_{t}[A(t)](\omega)$ needs to acquire a parabolic spectral phase $\widetilde{A}(\omega)\to\widetilde{A}(\omega)\exp[-i(f_{\mathrm{t}}/\omega_{0})\omega^{2}]$. Commonly, such an operation is realized directly by propagation in dispersive media \cite{karpinski_bandwidth_2017} or by employing a pulse stretcher/compressor.

\section{Temporal imaging using GEM and SSM}
\begin{figure}[!t]
\includegraphics[width=1\columnwidth]{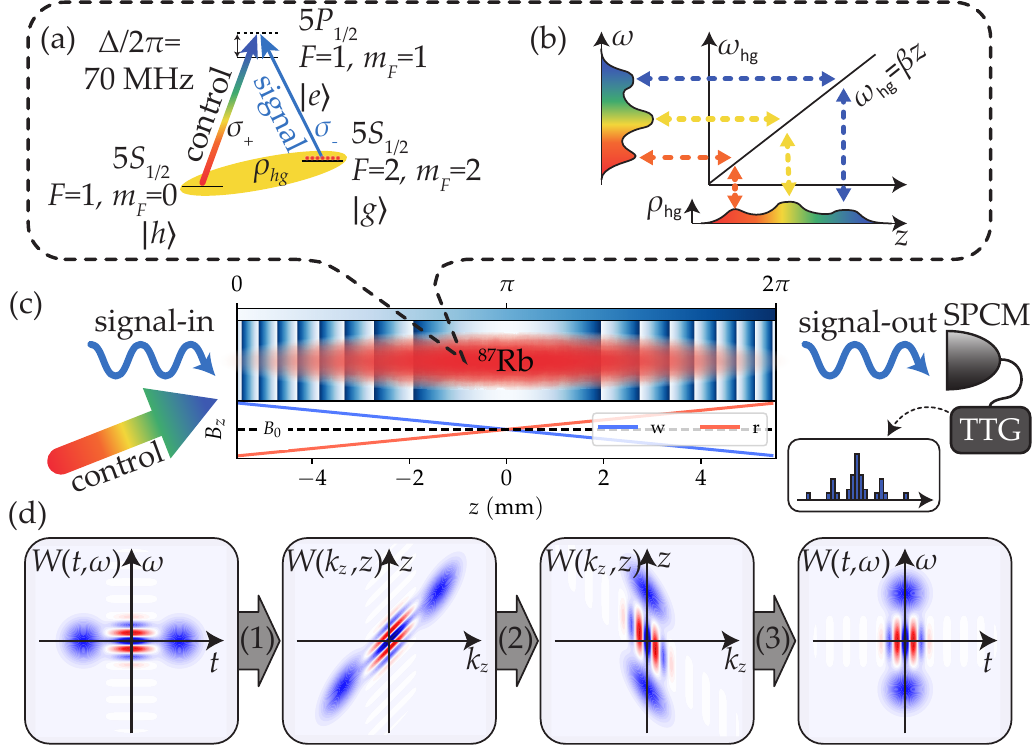}\caption{(a) Light-atom interface. Chirped control field simultaneously allows
mapping of the signal optical field onto the atomic coherence $\rho_{hg}$
and realizes the temporal lens. (b) Projection of signal spectral
components onto atomic coherence spatial components in GEM with Zeeman splitting gradient $\beta$. (c) During the writing process atoms are placed in a negative magnetic field gradient along the cloud (w). When the writing finishes, the spatial
phase of the atomic coherence is modulated with a parabolic Fresnel
profile which realizes a temporal equivalent of free-space propagation.
Finally, the gradient is switched to positive (r) and the coherence is converted back to light which is further registered
with Single-Photon-Counting-Module (SPCM) connected to the Time Tagger (TTG). (d) Evolution of the spectro-temporal Wigner function on
subsequent stages of far-field temporal imaging: (1) -- time-lens, (2) -- free-space propagation, (3) -- time-lens. The complete transformation effectively rotates the initial Wigner function of two pulses (equivalent to a Wigner function of a
cat state in phase space) by $\pi/2$ as given by Eq. \ref{eq:phase_space_rot}.}
\label{fig:setup}
\end{figure}

\begin{figure}[!t]
\includegraphics[width=1\columnwidth]{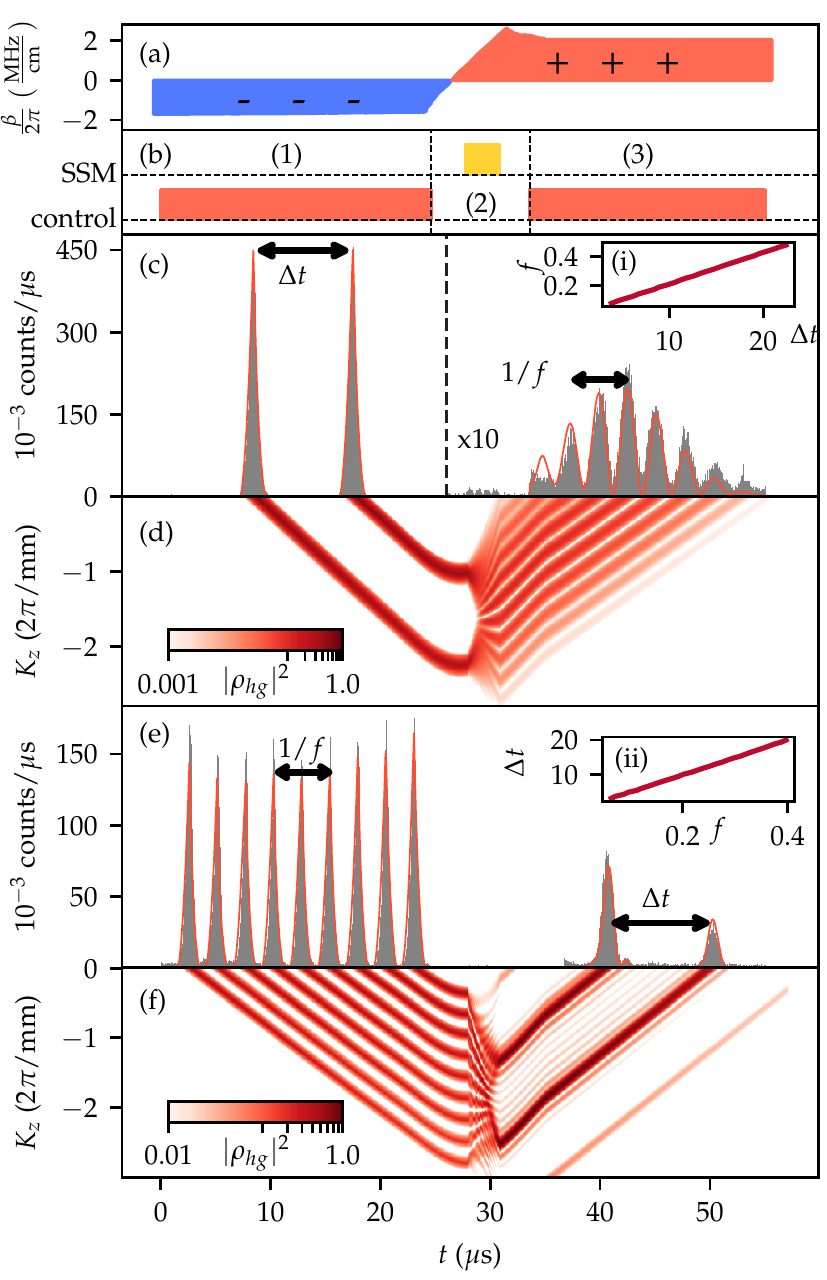}\caption{Experimental sequence for temporal imaging. (a) Time-trace of the Zeeman shift gradient $\beta$ used in GEM protocol, allowing two-directional mapping of signal frequencies to distinct positions in the atomic cloud. (b) Control field (red) and SSM (yellow) laser pulse sequence divided into three stages corresponding to lens-propagation-lens operations. The lens (1) is implemented during the GEM writing process by a chirped control field. (2) The 3 $\mu$s-long SSM laser pulse imprints a parabolic phase profile onto the stored atomic coherence which realizes the spectro-temporal free-space propagation. During this stage the magnetic field gradient is reversed, allowing re-mapping the coherence to light. (3) Finally, the control field is turned on and the coherence is read-out from the memory. 
Chirping the control field would implement the second lens. However, for simplicity, the control field is no longer chirped as the imposed phase would not be registered by the SPCM. 
(c,e) Example results, for two pulses
or a sine-wave as inputs, respectively. Gray bins represent single photon counts. Red line corresponds
to the numerical simulations. (d,f) Normalized modulus square of atomic coherence in Fourier space. The insets (i,ii) show experimentally obtained linear dependency of the time delay $\Delta t$ (in $\mu\mathrm{s}$) on the signal modulation frequency $f$ (in MHz) defined on panels (c,e).}
\label{fig:sin_fsin}
\end{figure}

In our technique we employ the atomic ensemble to process stored light and implement the temporal imaging operations during storage or light-atom mapping. The optical signal amplitude $A(t)$ is mapped onto the atomic coherence in a $\Lambda$ type system built of three atomic levels $|g\rangle$, $|h\rangle$ and $|e\rangle$ (see Fig. \ref{fig:setup} (a)). The mapping process employs a strong control field to make the atoms absorb the signal field and generate an atomic coherence $\rho_{hg}$ commonly called a spin-wave (SW). During the mapping (and re-mapping) process the atoms are kept in a magnetic field gradient, which constitutes the basis of the GEM \cite{Hosseini2009}, providing linearly changing Zeeman splitting between the $|h\rangle$ and $|e\rangle$ levels along the atomic cloud. This means that the signal-control two-photon interaction happens with a spatially dependent two-photon detuning $\delta$, and only atoms contained in a limited volume will interact efficiently with signal light of a specific frequency. Therefore, distinct spectral components of signal light $\widetilde{A}(\omega)$ are mapped onto different spatial components of the atomic coherence $\rho_{hg}(z)\propto\widetilde{A}(\beta z)$  \cite{Hosseini2009,Sparkes2013} (see Fig. \ref{fig:setup} (b)) and vice versa, where $\beta$ denotes the gradient of the Zeeman shift along the propagation ($z$) axis. 

The temporal equivalent of free space propagation is realized thanks to this spectro-spatial mapping, characteristic of the GEM. Spatially-resolved phase modulation of a SW stored in the memory is equivalent to imposing a spectral phase profile onto the signal. Thus, by imposing onto the SWs a parabolic spatial phase $\exp[-if_{\mathrm{t}}/(2\omega_{0})\beta^{2}z^{2}]$ we implement the temporal analog of free-space propagation. This operation is implemented using the spatially-resolved ac-Stark shift induced by an additional far-detuned and spatially-shaped laser beam -- a technique we call spatial spin-wave modulation (SSM) \cite{Leszczynski2018,Parniak2019,mazelanik_coherent_2019,Lipka2019}.

To make the time-lens we utilize the fact that the SW is created in coherent two-photon absorption, thus it reflects the temporal phase difference between the control and signal field. This means, that chirping the signal field is equivalent to chirping the control field, as only the two-photon detuning $\delta$ is crucial here. Hence, by changing the control field frequency, we make the two-photon detuning linearly time dependent $\delta=\alpha t$ and impose the desired quadratic phase during the interaction time. This way the time-lens is realized at the light-SW mapping stage without affecting the signal field directly. Yet, as we chose the single-photon detuning $\Delta\gg\delta$ residual modulation of the coupling efficiency is negligible as $\Delta+\alpha t\approx\Delta$.

Finally, sending a signal field $A(t)$ through a lens-propagation-lens temporal imaging system, the output amplitude is proportional to $\widetilde{A}\left(\alpha t\right)$. In practice, however,
the finite size of the atomic cloud must be taken into account making
the output amplitude proportional to $\left(\left[\widetilde{A}(\alpha t)\exp[-i(\alpha/2)t^{2}]\right]*\zeta(t)*\zeta(t)\right)\exp[i(\alpha/2)t^{2}]$,
where $\zeta(t)=\mathcal{F}_{\omega}[\eta_{0}(\omega)](t)$ is the
Fourier transform of the inhomogeneously broadened absorption efficiency
spectrum $\eta_{0}(\omega)$ determined by the atomic density distribution
and field gradient $\beta$, and $*$ symbolizes convolution.

In a typical regime of operation we select the chirp $\alpha \ll(\beta L)^{2}$
to always contain the entire spectrum of the pulse within the atomic
absorption bandwidth $\mathcal{B}\approx\beta L$. The resolution
in this regime is limited by the decoherence of spin-waves caused
by the control beam of the light-atom interface and is given by the inverse
of the atomic coherence lifetime $\delta\omega/2\pi=0.78/\tau$ (see
\sm  for derivation of the prefactor), where $1/\tau=\Gamma\Omega^{2}/(4\Delta^{2}+\Gamma^{2})$
and $\Gamma$ is the decay rate of the $|e\rangle$ state and $\Omega$
is the Rabi frequency at the $|h\rangle\rightarrow|e\rangle$ transition.

\section{Experiment}
The core of our setup is a GEM based on a cold $^{87}$Rb atomic ensemble trapped in a magneto-optical trap (MOT) over 1-cm-long pencil-shaped volume. After the MOT release, all atoms are optically pumped to the $|g\rangle=5\mathrm{S}_{1/2}$, $F=2$, $m_{F}=2$ state. The ensemble optical depth reaches $\mathrm{{OD}}\sim70$ at the $|g\rangle\rightarrow|e\rangle=5P_{1/2}$, $F=1$, $m_{F}=1$ transition. As depicted in Fig. \ref{fig:setup}(a), we employ the $\Lambda$ system to couple light signal and atomic coherence (spin-waves). The interface consists of a $\sigma_{+}$ polarized strong control laser blue-detuned by $\Delta=2\pi\times70$ MHz from the $|h\rangle=5S_{1/2}$, $F=1$, $m_{F}=0\rightarrow|e\rangle$ transition and a weak $\sigma_{-}$ polarized signal laser at the $|g\rangle\rightarrow|e\rangle$ transition, approximately at the two-photon resonance $\delta\approx0$. The gradient $\beta$ of the Zeeman splitting along the $z$ axis during signal-to-coherence conversion is generated by two rounded-square shaped coils connected in opposite configuration (see \sm for details).
The SSM scheme facilitates manipulation of the spatial phase of stored spin-waves via off-resonant ac-Stark shift by illuminating the atomic cloud with a spatially shaped strong $\pi$-polarized beam, 1 GHz blue-detuned from the $5S_{1/2}$, $F=1\rightarrow5P_{3/2}$ transition. The signal emitted in $|g\rangle\rightarrow|e\rangle$ transition is filtered using Wollaston polarizer and an optically-pumped atomic filter, to be finally registered on a Single Photon Counting Module (SPCM). We finally register only $0.023$ noise counts on average per $\tau$-long detection window (see \sm).

In Fig. \ref{fig:sin_fsin} we present exemplary measurements performed with our setup. Panel (a) shows the time trace of the Zeeman shift gradient set initially to $\beta=-2\pi\times1.7$ MHz/cm. In panel (b) we provide the control and SSM laser sequence divided into stages corresponding to subsequent implementations of lens-propagation-lens operations. (1) First, a strong control field (Rabi frequency $\Omega=2\pi\times4.7$ MHz) is used to map a weak ($\bar{n}=2.8$) signal pulse with temporal amplitude $A(t)$ to the atomic coherence. The control beam is chirped with an acousto-optic modulator (AOM) to have a time-dependent frequency of $\omega(t)=\omega_{0}+\alpha t$, with $\alpha=2\pi\times0.04$ MHz/$\mu$s. This implements a time-lens with focal length $f_{\mathrm{t}}=9.6\times10^3$ s. (2) Next, within 7 $\mu$s the gradient $\beta$ is switched to opposite value and a parabolic Fresnel phase profile $\exp[-i\beta^{2}/(2\alpha)z^{2}]$ (as depicted in Fig. \ref{fig:setup}(c)) is imprinted onto stored atomic coherence by the 3~$\mu$s long SSM laser pulse. The linear gradient of magnetic field only shifts atomic coherence in the Fourier domain, therefore phase modulation can be done simultaneously with $\beta$ reversing. (3) Finally, the control field is turned on and the coherence is converted back to light. For simplicity, during GEM readout the control field is no longer chirped as the imposed phase would not be registered by the the SPCM (see \sm for details). 

Figure \ref{fig:sin_fsin}(c-f) portrays experimental results for two types of the input signal: (c,d) two peaks and (e,f) sine-wave-like waveform. 
Red solid lines corresponds to the full light-atoms interaction simulation (see \sm for details). The density maps (d,f) below each time trace (c,e) show simulated evolution of the atomic coherence during the experiment. For both input signal shapes the measured efficiency amounts to about 7\%. The insets (i,ii) show experimentally obtained linear relationship between time delay $\Delta t$ and the signal modulation (temporal fringes) frequency $f$ defined in panels (c,e).

We attribute the residual mismatch between experimental results and theoretical predictions to imperfect linearity of magnetic field gradient, decoherence caused by ac-Stark modulation and a simplification of the atomic density distribution in the simulation. However, for both exemplary measurements we can still observe a good agreement with the theory. Notably, the simulations use independently calibrated parameters, with only the input photon number adjusted for the specific measurements.

\begin{figure}[!t]
\includegraphics[width=1\columnwidth]{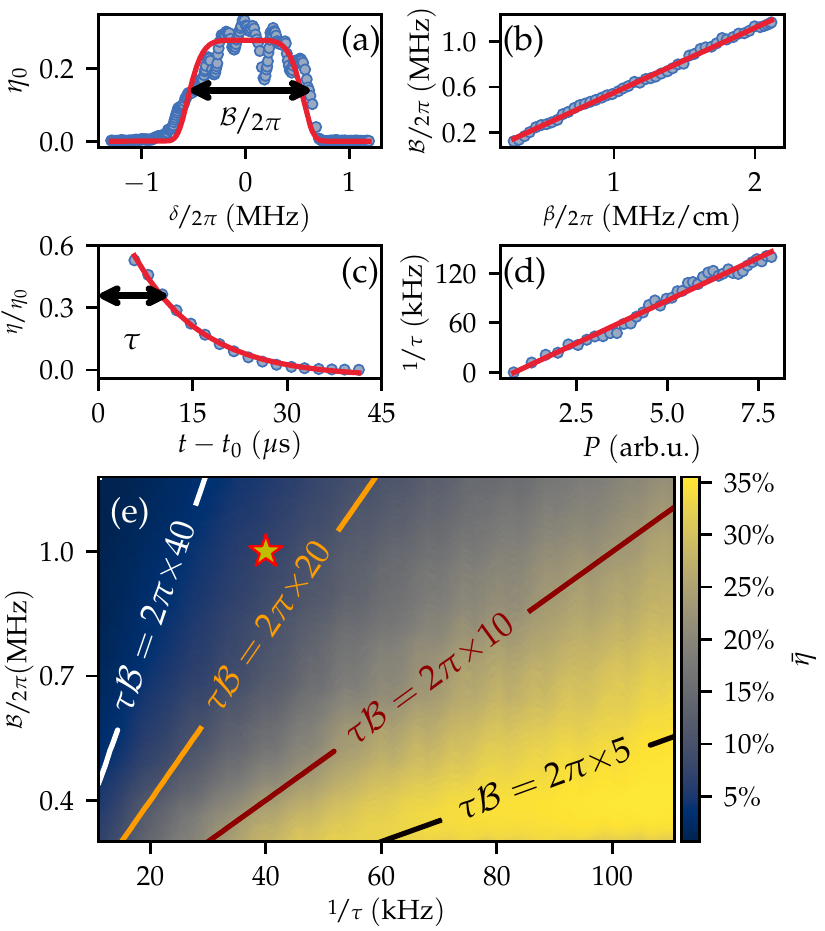}\caption{Characterization and tuning of bandwidth and resolution. (a) Efficiency
spectral profile $\eta_{0}(\omega)$ as a function of the two-photon
detuning $\delta=\omega-\omega_{0}$ for a chosen time-bandwidth product
$\tau\mathcal{B}=2\pi\times13$ with bandwidth $\mathcal{B}$ defined as FWHM
of $\eta_{0}(\omega)$. The red line corresponds to a super-Gaussian approximation of the atom concentration used in the simulation. (b) Dependence of the bandwidth $\mathcal{B}$
as a function of the Zeeman splitting gradient $\beta$. Red line is a linear fit to the data. (c) Time evolution
of the GEM efficiency due to incoherent scattering caused by the coupling
field. The characteristic decay time $\tau$ obtained from exponential fit (red line) limits the effective
resolution $\delta\omega/2\pi=0.78/\tau$ (here $\tau=10\ \mu\mathrm{s}$).
(d) Dependence of $1/\tau$ as a function of the coupling field
power $P\propto|\Omega|^{2}$, along with linear fit (red line). (e) Calculated map of the average efficiency $\bar{\eta}$
for varying bandwidth $\mathcal{B}$ and decay time $\tau$. The efficiency
for a given time-bandwidth product $\tau\mathcal{B}$ is approximately
constant as expected. The star indicates the point of operation where the exemplary measurements (Fig. \ref{fig:sin_fsin}) were performed. \label{fig:TB}}
\end{figure}
\section{Time-bandwidth characterization}
Figures of merit characterizing our device are bandwidth, resolution
and efficiency. Those parameters are related by a formula for GEM
efficiency \cite{Sparkes2013} which for atoms uniformly distributed
over the length $L$ becomes $\omega$-independent and can be approximated
as 
\begin{equation}
\eta_{0}=\left[1-\exp\left(-2\pi\frac{\mathrm{OD}}{\tau\mathcal{B}}\right)\right]^{2},\label{eq:eta0}
\end{equation}
where OD is the optical depth of the ensemble for $\Delta=0$. Equation
\ref{eq:eta0} indicates that increased bandwidth or resolution results
in a drop in efficiency. In a realistic scenario atoms are non-uniformly
distributed over the cloud and thus different spectral components
of the input field experience different values of OD, especially at
the edges of the atomic cloud. This makes the efficiency $\eta_{0}$
frequency-dependent and leads to an operational definition of the
bandwidth $\mathcal{B}$ as the FWHM of the $\eta_{0}(\omega)$ profile
as depicted in Fig. \ref{fig:TB}(a). Additionally, due to the decoherence
induced by the coupling field during the write (and read) process,
the efficiency decays exponentially in time: $\eta=\eta_{0}\Theta(t)\exp(-t/\tau)$,
as illustrated in Fig. \ref{fig:TB}(c). Therefore, to account for
spectro-temporal dependencies, we introduce a time-frequency averaged
efficiency: 
\begin{equation}
\bar{\eta}=\frac{1}{2\tau\mathcal{B}}\int_{-\mathcal{B}/2}^{\mathcal{B}/2}\int_{0}^{2\tau}\eta(t,\omega)\mathrm{d}t\mathrm{d}\omega=\frac{e^2-1}{2e^2\mathcal{B}}\int_{-\mathcal{B}/2}^{\mathcal{B}/2}\eta_{0}(\omega)\mathrm{d}\omega.
\end{equation}

Figure \ref{fig:TB}(e) illustrates measured values of $\bar{\eta}$
for different $\mathcal{B}$ and $\tau$. The map is built from separate measurements of $\eta_0(\omega)$ (Fig. \ref{fig:TB}(a)) and $\tau$ (Fig. \ref{fig:TB}(c)) for different Zeeman shift gradient $\beta$ (Fig. \ref{fig:TB}(b)) and control laser power $P\propto|\Omega|^2$ (Fig. \ref{fig:TB}(d)). The parameters extracted from Fig. \ref{fig:TB}(a-d) are then combined to give value of $\bar{\eta}$ for each $(\tau(P),\mathcal{B}(\beta))$ point. As expected from Eq. \ref{eq:eta0}, we see a clear trade-off between the time-bandwidth product $\tau\mathcal{B}$
and the average efficiency $\bar{\eta}$. Conversely, requiring a
higher number of distinguishable frequency (or time) bins leads to
a lower efficiency. Yet, with $\sim20$\% mean efficiency we obtain
$\tau\mathcal{B}=2\pi\times10$ that simultaneously yields 100 kHz resolution
and 1 MHz bandwidth. One may also choose to maximize $\tau\mathcal{B}$ to reach $2\pi\times40$, with mean efficiency $\sim4\%$. Notably, as the efficiency $\eta_0$
saturates for large OD, for systems with ultra-high optical depth the time-bandwidth product could reach
significantly higher values while maintaining near-unity efficiency for many~bins.

\section{Conclusions}
In summary, we have introduced and experimentally demonstrated a novel
high-resolution (ca. 20 kHz) far-field imaging method suitable for
narrow-band atomic photon sources -- a region previously unattainable.
The device may also serve as a single-photon-level ultra-precise spectrometer
for atomic emission, enabling characterization of spectro-temporal
high-dimensional entanglement generated with atoms. In general, while
temporal domain characterization and manipulation at the single-photon
level is already indispensable in numerous quantum information processing
tasks, quantum networks architectures and metrology, our device will
allow those techniques to enter the ultra-narrow bandwidth domain.
Our method utilizes a multi-mode gradient echo memory (GEM) along
a recently developed processing technique -- spatial spin-wave modulator
(SSM) \cite{Parniak2019,mazelanik_coherent_2019,Lipka2019} -- enabling
nearly arbitrary manipulations on light states stored in GEM. We envisage that with improvement of the magnetic field gradient the GEM bandwidth can reach dozens of MHz opening new ranges of applications such as solid-state quantum memories \cite{Hedges2010} and color centers \cite{Jeong2019}. Furthermore,
our approach utilizes a quantum memory previously demonstrated \cite{Parniak2017,mazelanik_coherent_2019}
to operate with quantum states of light, and maintains the ultra low
level of noise, creating novel possibilities in temporal and spectral
processing of narrow-band atomic-emission quantum states of light.
Our technique applied to systems with higher absorption bandwidth
\cite{Saglamyurek2018} or optical depth \cite{Cho2016} can bridge
the bandwidth gap to enable hybrid solid-state--atomic quantum networks
operating using the full temporal-spectral degree of freedom. 

\paragraph{Funding.}

MNiSW (DI2016 014846, DI2018 010848); National Science Centre (2016/21/B/ST2/02559,
2017/25/N/ST2/01163, 2017/25/N/ST2/00713); 
Foundation for Polish Science (MAB 4/2017 "Quantum Optical Technologies"); Office of Naval Research (N62909-19-1-2127).

\paragraph{Acknowledgments.}

We thank K. Banaszek for the generous support and M. Jachura for insightful discussion. M.P. was supported
by the Foundation for Polish Science via the START scholarship. A.L. and M.M. contributed equally
to this work. The "Quantum Optical Technologies ” project is carried
out within the International Research Agendas programme
of the Foundation for Polish Science co-financed
by the European Union under the European Regional
Development Fund.

M.M. and A.L. contributed equally to this work.



\bibliographystyle{apsrev4-1}
\bibliography{swft}

\begin{thebibliography}{55}%
\makeatletter
\providecommand \@ifxundefined [1]{%
 \@ifx{#1\undefined}
}%
\providecommand \@ifnum [1]{%
 \ifnum #1\expandafter \@firstoftwo
 \else \expandafter \@secondoftwo
 \fi
}%
\providecommand \@ifx [1]{%
 \ifx #1\expandafter \@firstoftwo
 \else \expandafter \@secondoftwo
 \fi
}%
\providecommand \natexlab [1]{#1}%
\providecommand \enquote  [1]{``#1''}%
\providecommand \bibnamefont  [1]{#1}%
\providecommand \bibfnamefont [1]{#1}%
\providecommand \citenamefont [1]{#1}%
\providecommand \href@noop [0]{\@secondoftwo}%
\providecommand \href [0]{\begingroup \@sanitize@url \@href}%
\providecommand \@href[1]{\@@startlink{#1}\@@href}%
\providecommand \@@href[1]{\endgroup#1\@@endlink}%
\providecommand \@sanitize@url [0]{\catcode `\\12\catcode `\$12\catcode
  `\&12\catcode `\#12\catcode `\^12\catcode `\_12\catcode `\%12\relax}%
\providecommand \@@startlink[1]{}%
\providecommand \@@endlink[0]{}%
\providecommand \url  [0]{\begingroup\@sanitize@url \@url }%
\providecommand \@url [1]{\endgroup\@href {#1}{\urlprefix }}%
\providecommand \urlprefix  [0]{URL }%
\providecommand \Eprint [0]{\href }%
\providecommand \doibase [0]{http://dx.doi.org/}%
\providecommand \selectlanguage [0]{\@gobble}%
\providecommand \bibinfo  [0]{\@secondoftwo}%
\providecommand \bibfield  [0]{\@secondoftwo}%
\providecommand \translation [1]{[#1]}%
\providecommand \BibitemOpen [0]{}%
\providecommand \bibitemStop [0]{}%
\providecommand \bibitemNoStop [0]{.\EOS\space}%
\providecommand \EOS [0]{\spacefactor3000\relax}%
\providecommand \BibitemShut  [1]{\csname bibitem#1\endcsname}%
\let\auto@bib@innerbib\@empty
\bibitem [{\citenamefont {Humphreys}\ \emph {et~al.}(2014)\citenamefont
  {Humphreys}, \citenamefont {Kolthammer}, \citenamefont {Nunn}, \citenamefont
  {Barbieri}, \citenamefont {Datta},\ and\ \citenamefont
  {Walmsley}}]{Humphreys2014}%
  \BibitemOpen
  \bibfield  {author} {\bibinfo {author} {\bibfnamefont {P.~C.}\ \bibnamefont
  {Humphreys}}, \bibinfo {author} {\bibfnamefont {W.~S.}\ \bibnamefont
  {Kolthammer}}, \bibinfo {author} {\bibfnamefont {J.}~\bibnamefont {Nunn}},
  \bibinfo {author} {\bibfnamefont {M.}~\bibnamefont {Barbieri}}, \bibinfo
  {author} {\bibfnamefont {A.}~\bibnamefont {Datta}}, \ and\ \bibinfo {author}
  {\bibfnamefont {I.~A.}\ \bibnamefont {Walmsley}},\ }\href {\doibase
  10.1103/PhysRevLett.113.130502} {\bibfield  {journal} {\bibinfo  {journal}
  {Phys. Rev. Lett.}\ }\textbf {\bibinfo {volume} {113}},\ \bibinfo {pages}
  {130502} (\bibinfo {year} {2014})},\ \Eprint {http://arxiv.org/abs/1405.5361}
  {arXiv:1405.5361} \BibitemShut {NoStop}%
\bibitem [{\citenamefont {Brecht}\ \emph {et~al.}(2015)\citenamefont {Brecht},
  \citenamefont {Reddy}, \citenamefont {Silberhorn},\ and\ \citenamefont
  {Raymer}}]{Brecht2015}%
  \BibitemOpen
  \bibfield  {author} {\bibinfo {author} {\bibfnamefont {B.}~\bibnamefont
  {Brecht}}, \bibinfo {author} {\bibfnamefont {D.~V.}\ \bibnamefont {Reddy}},
  \bibinfo {author} {\bibfnamefont {C.}~\bibnamefont {Silberhorn}}, \ and\
  \bibinfo {author} {\bibfnamefont {M.~G.}\ \bibnamefont {Raymer}},\ }\href
  {\doibase 10.1103/PhysRevX.5.041017} {\bibfield  {journal} {\bibinfo
  {journal} {Physical Review X}\ }\textbf {\bibinfo {volume} {5}},\ \bibinfo
  {pages} {041017} (\bibinfo {year} {2015})},\ \Eprint
  {http://arxiv.org/abs/1504.06251} {arXiv:1504.06251} \BibitemShut {NoStop}%
\bibitem [{\citenamefont {Reimer}\ \emph {et~al.}(2016)\citenamefont {Reimer},
  \citenamefont {Kues}, \citenamefont {Roztocki}, \citenamefont {Wetzel},
  \citenamefont {Grazioso}, \citenamefont {Little}, \citenamefont {Chu},
  \citenamefont {Johnston}, \citenamefont {Bromberg}, \citenamefont {Caspani},
  \citenamefont {Moss},\ and\ \citenamefont {Morandotti}}]{Reimer2016}%
  \BibitemOpen
  \bibfield  {author} {\bibinfo {author} {\bibfnamefont {C.}~\bibnamefont
  {Reimer}}, \bibinfo {author} {\bibfnamefont {M.}~\bibnamefont {Kues}},
  \bibinfo {author} {\bibfnamefont {P.}~\bibnamefont {Roztocki}}, \bibinfo
  {author} {\bibfnamefont {B.}~\bibnamefont {Wetzel}}, \bibinfo {author}
  {\bibfnamefont {F.}~\bibnamefont {Grazioso}}, \bibinfo {author}
  {\bibfnamefont {B.~E.}\ \bibnamefont {Little}}, \bibinfo {author}
  {\bibfnamefont {S.~T.}\ \bibnamefont {Chu}}, \bibinfo {author} {\bibfnamefont
  {T.}~\bibnamefont {Johnston}}, \bibinfo {author} {\bibfnamefont
  {Y.}~\bibnamefont {Bromberg}}, \bibinfo {author} {\bibfnamefont
  {L.}~\bibnamefont {Caspani}}, \bibinfo {author} {\bibfnamefont {D.~J.}\
  \bibnamefont {Moss}}, \ and\ \bibinfo {author} {\bibfnamefont
  {R.}~\bibnamefont {Morandotti}},\ }\href {\doibase 10.1126/science.aad8532}
  {\bibfield  {journal} {\bibinfo  {journal} {Science}\ }\textbf {\bibinfo
  {volume} {351}},\ \bibinfo {pages} {1176} (\bibinfo {year}
  {2016})}\BibitemShut {NoStop}%
\bibitem [{\citenamefont {Lu}\ \emph {et~al.}(2019)\citenamefont {Lu},
  \citenamefont {Lukens}, \citenamefont {Williams}, \citenamefont {Imany},
  \citenamefont {Peters}, \citenamefont {Weiner},\ and\ \citenamefont
  {Lougovski}}]{Lu2019}%
  \BibitemOpen
  \bibfield  {author} {\bibinfo {author} {\bibfnamefont {H.-H.}\ \bibnamefont
  {Lu}}, \bibinfo {author} {\bibfnamefont {J.~M.}\ \bibnamefont {Lukens}},
  \bibinfo {author} {\bibfnamefont {B.~P.}\ \bibnamefont {Williams}}, \bibinfo
  {author} {\bibfnamefont {P.}~\bibnamefont {Imany}}, \bibinfo {author}
  {\bibfnamefont {N.~A.}\ \bibnamefont {Peters}}, \bibinfo {author}
  {\bibfnamefont {A.~M.}\ \bibnamefont {Weiner}}, \ and\ \bibinfo {author}
  {\bibfnamefont {P.}~\bibnamefont {Lougovski}},\ }\href {\doibase
  10.1038/s41534-019-0137-z} {\bibfield  {journal} {\bibinfo  {journal} {npj
  Quantum Inf.}\ }\textbf {\bibinfo {volume} {5}},\ \bibinfo {pages} {24}
  (\bibinfo {year} {2019})},\ \Eprint {http://arxiv.org/abs/1809.05072}
  {arXiv:1809.05072} \BibitemShut {NoStop}%
\bibitem [{\citenamefont {Pu}\ \emph {et~al.}(2017)\citenamefont {Pu},
  \citenamefont {Jiang}, \citenamefont {Chang}, \citenamefont {Yang},
  \citenamefont {Li},\ and\ \citenamefont {Duan}}]{Pu2017}%
  \BibitemOpen
  \bibfield  {author} {\bibinfo {author} {\bibfnamefont {Y.~F.}\ \bibnamefont
  {Pu}}, \bibinfo {author} {\bibfnamefont {N.}~\bibnamefont {Jiang}}, \bibinfo
  {author} {\bibfnamefont {W.}~\bibnamefont {Chang}}, \bibinfo {author}
  {\bibfnamefont {H.~X.}\ \bibnamefont {Yang}}, \bibinfo {author}
  {\bibfnamefont {C.}~\bibnamefont {Li}}, \ and\ \bibinfo {author}
  {\bibfnamefont {L.~M.}\ \bibnamefont {Duan}},\ }\href {\doibase
  10.1038/ncomms15359} {\bibfield  {journal} {\bibinfo  {journal} {Nat.
  Commun.}\ }\textbf {\bibinfo {volume} {8}},\ \bibinfo {pages} {15359}
  (\bibinfo {year} {2017})},\ \Eprint {http://arxiv.org/abs/1707.07267}
  {arXiv:1707.07267} \BibitemShut {NoStop}%
\bibitem [{\citenamefont {Parniak}\ \emph {et~al.}(2017)\citenamefont
  {Parniak}, \citenamefont {Dąbrowski}, \citenamefont {Mazelanik},
  \citenamefont {Leszczy{\'{n}}ski}, \citenamefont {Lipka},\ and\ \citenamefont
  {Wasilewski}}]{Parniak2017}%
  \BibitemOpen
  \bibfield  {author} {\bibinfo {author} {\bibfnamefont {M.}~\bibnamefont
  {Parniak}}, \bibinfo {author} {\bibfnamefont {M.}~\bibnamefont {Dąbrowski}},
  \bibinfo {author} {\bibfnamefont {M.}~\bibnamefont {Mazelanik}}, \bibinfo
  {author} {\bibfnamefont {A.}~\bibnamefont {Leszczy{\'{n}}ski}}, \bibinfo
  {author} {\bibfnamefont {M.}~\bibnamefont {Lipka}}, \ and\ \bibinfo {author}
  {\bibfnamefont {W.}~\bibnamefont {Wasilewski}},\ }\href {\doibase
  10.1038/s41467-017-02366-7} {\bibfield  {journal} {\bibinfo  {journal} {Nat.
  Commun.}\ }\textbf {\bibinfo {volume} {8}},\ \bibinfo {pages} {2140}
  (\bibinfo {year} {2017})},\ \Eprint {http://arxiv.org/abs/1706.04426}
  {arXiv:1706.04426} \BibitemShut {NoStop}%
\bibitem [{\citenamefont {Mazelanik}\ \emph {et~al.}(2019)\citenamefont
  {Mazelanik}, \citenamefont {Parniak}, \citenamefont {Leszczy{\'{n}}ski},
  \citenamefont {Lipka},\ and\ \citenamefont
  {Wasilewski}}]{mazelanik_coherent_2019}%
  \BibitemOpen
  \bibfield  {author} {\bibinfo {author} {\bibfnamefont {M.}~\bibnamefont
  {Mazelanik}}, \bibinfo {author} {\bibfnamefont {M.}~\bibnamefont {Parniak}},
  \bibinfo {author} {\bibfnamefont {A.}~\bibnamefont {Leszczy{\'{n}}ski}},
  \bibinfo {author} {\bibfnamefont {M.}~\bibnamefont {Lipka}}, \ and\ \bibinfo
  {author} {\bibfnamefont {W.}~\bibnamefont {Wasilewski}},\ }\href {\doibase
  10.1038/s41534-019-0136-0} {\bibfield  {journal} {\bibinfo  {journal} {npj
  Quantum Inf.}\ }\textbf {\bibinfo {volume} {5}},\ \bibinfo {pages} {22}
  (\bibinfo {year} {2019})},\ \Eprint {http://arxiv.org/abs/1808.00927}
  {arXiv:1808.00927} \BibitemShut {NoStop}%
\bibitem [{\citenamefont {Seri}\ \emph {et~al.}(2019)\citenamefont {Seri},
  \citenamefont {Lago-Rivera}, \citenamefont {Lenhard}, \citenamefont
  {Corrielli}, \citenamefont {Osellame}, \citenamefont {Mazzera},\ and\
  \citenamefont {de~Riedmatten}}]{Seri2019}%
  \BibitemOpen
  \bibfield  {author} {\bibinfo {author} {\bibfnamefont {A.}~\bibnamefont
  {Seri}}, \bibinfo {author} {\bibfnamefont {D.}~\bibnamefont {Lago-Rivera}},
  \bibinfo {author} {\bibfnamefont {A.}~\bibnamefont {Lenhard}}, \bibinfo
  {author} {\bibfnamefont {G.}~\bibnamefont {Corrielli}}, \bibinfo {author}
  {\bibfnamefont {R.}~\bibnamefont {Osellame}}, \bibinfo {author}
  {\bibfnamefont {M.}~\bibnamefont {Mazzera}}, \ and\ \bibinfo {author}
  {\bibfnamefont {H.}~\bibnamefont {de~Riedmatten}},\ }\href {\doibase
  10.1103/physrevlett.123.080502} {\bibfield  {journal} {\bibinfo  {journal}
  {Phys. Rev. Lett.}\ }\textbf {\bibinfo {volume} {123}},\ \bibinfo {pages}
  {080502} (\bibinfo {year} {2019})},\ \Eprint
  {http://arxiv.org/abs/1902.06657} {arXiv:1902.06657} \BibitemShut {NoStop}%
\bibitem [{\citenamefont {Cho}\ \emph {et~al.}(2016)\citenamefont {Cho},
  \citenamefont {Campbell}, \citenamefont {Everett}, \citenamefont {Bernu},
  \citenamefont {Higginbottom}, \citenamefont {Cao}, \citenamefont {Geng},
  \citenamefont {Robins}, \citenamefont {Lam},\ and\ \citenamefont
  {Buchler}}]{Cho2016}%
  \BibitemOpen
  \bibfield  {author} {\bibinfo {author} {\bibfnamefont {Y.-W.}\ \bibnamefont
  {Cho}}, \bibinfo {author} {\bibfnamefont {G.~T.}\ \bibnamefont {Campbell}},
  \bibinfo {author} {\bibfnamefont {J.~L.}\ \bibnamefont {Everett}}, \bibinfo
  {author} {\bibfnamefont {J.}~\bibnamefont {Bernu}}, \bibinfo {author}
  {\bibfnamefont {D.~B.}\ \bibnamefont {Higginbottom}}, \bibinfo {author}
  {\bibfnamefont {M.~T.}\ \bibnamefont {Cao}}, \bibinfo {author} {\bibfnamefont
  {J.}~\bibnamefont {Geng}}, \bibinfo {author} {\bibfnamefont {N.~P.}\
  \bibnamefont {Robins}}, \bibinfo {author} {\bibfnamefont {P.~K.}\
  \bibnamefont {Lam}}, \ and\ \bibinfo {author} {\bibfnamefont {B.~C.}\
  \bibnamefont {Buchler}},\ }\href {\doibase 10.1364/optica.3.000100}
  {\bibfield  {journal} {\bibinfo  {journal} {Optica}\ }\textbf {\bibinfo
  {volume} {3}},\ \bibinfo {pages} {100} (\bibinfo {year} {2016})},\ \Eprint
  {http://arxiv.org/abs/1601.04267} {arXiv:1601.04267} \BibitemShut {NoStop}%
\bibitem [{\citenamefont {Bao}\ \emph {et~al.}(2012)\citenamefont {Bao},
  \citenamefont {Reingruber}, \citenamefont {Dietrich}, \citenamefont {Rui},
  \citenamefont {D{\"{u}}ck}, \citenamefont {Strassel}, \citenamefont {Li},
  \citenamefont {Liu}, \citenamefont {Zhao},\ and\ \citenamefont
  {Pan}}]{Bao2012}%
  \BibitemOpen
  \bibfield  {author} {\bibinfo {author} {\bibfnamefont {X.~H.}\ \bibnamefont
  {Bao}}, \bibinfo {author} {\bibfnamefont {A.}~\bibnamefont {Reingruber}},
  \bibinfo {author} {\bibfnamefont {P.}~\bibnamefont {Dietrich}}, \bibinfo
  {author} {\bibfnamefont {J.}~\bibnamefont {Rui}}, \bibinfo {author}
  {\bibfnamefont {A.}~\bibnamefont {D{\"{u}}ck}}, \bibinfo {author}
  {\bibfnamefont {T.}~\bibnamefont {Strassel}}, \bibinfo {author}
  {\bibfnamefont {L.}~\bibnamefont {Li}}, \bibinfo {author} {\bibfnamefont
  {N.~L.}\ \bibnamefont {Liu}}, \bibinfo {author} {\bibfnamefont
  {B.}~\bibnamefont {Zhao}}, \ and\ \bibinfo {author} {\bibfnamefont {J.~W.}\
  \bibnamefont {Pan}},\ }\href {\doibase 10.1038/nphys2324} {\bibfield
  {journal} {\bibinfo  {journal} {Nat. Phys.}\ }\textbf {\bibinfo {volume}
  {8}},\ \bibinfo {pages} {517} (\bibinfo {year} {2012})},\ \Eprint
  {http://arxiv.org/abs/1207.2894} {arXiv:1207.2894} \BibitemShut {NoStop}%
\bibitem [{\citenamefont {Kolner}\ and\ \citenamefont
  {Nazarathy}(1989)}]{kolner_temporal_1989}%
  \BibitemOpen
  \bibfield  {author} {\bibinfo {author} {\bibfnamefont {B.~H.}\ \bibnamefont
  {Kolner}}\ and\ \bibinfo {author} {\bibfnamefont {M.}~\bibnamefont
  {Nazarathy}},\ }\href {\doibase 10.1364/OL.14.000630} {\bibfield  {journal}
  {\bibinfo  {journal} {Opt. Lett.}\ }\textbf {\bibinfo {volume} {14}},\
  \bibinfo {pages} {630} (\bibinfo {year} {1989})}\BibitemShut {NoStop}%
\bibitem [{\citenamefont {Zhu}\ \emph {et~al.}(2013)\citenamefont {Zhu},
  \citenamefont {Kim},\ and\ \citenamefont
  {Gauthier}}]{zhu_aberration-corrected_2013}%
  \BibitemOpen
  \bibfield  {author} {\bibinfo {author} {\bibfnamefont {Y.}~\bibnamefont
  {Zhu}}, \bibinfo {author} {\bibfnamefont {J.}~\bibnamefont {Kim}}, \ and\
  \bibinfo {author} {\bibfnamefont {D.~J.}\ \bibnamefont {Gauthier}},\ }\href
  {\doibase 10.1103/PhysRevA.87.043808} {\bibfield  {journal} {\bibinfo
  {journal} {Phys. Rev. A}\ }\textbf {\bibinfo {volume} {87}},\ \bibinfo
  {pages} {43808} (\bibinfo {year} {2013})}\BibitemShut {NoStop}%
\bibitem [{\citenamefont {Patera}\ \emph {et~al.}(2018)\citenamefont {Patera},
  \citenamefont {Horoshko},\ and\ \citenamefont
  {Kolobov}}]{patera_space-time_2018}%
  \BibitemOpen
  \bibfield  {author} {\bibinfo {author} {\bibfnamefont {G.}~\bibnamefont
  {Patera}}, \bibinfo {author} {\bibfnamefont {D.~B.}\ \bibnamefont
  {Horoshko}}, \ and\ \bibinfo {author} {\bibfnamefont {M.~I.}\ \bibnamefont
  {Kolobov}},\ }\href {\doibase 10.1103/PhysRevA.98.053815} {\bibfield
  {journal} {\bibinfo  {journal} {Phys. Rev. A}\ }\textbf {\bibinfo {volume}
  {98}},\ \bibinfo {pages} {053815} (\bibinfo {year} {2018})},\ \Eprint
  {http://arxiv.org/abs/1806.11181} {arXiv:1806.11181} \BibitemShut {NoStop}%
\bibitem [{\citenamefont {Kolner}(1988)}]{kolner_active_1988}%
  \BibitemOpen
  \bibfield  {author} {\bibinfo {author} {\bibfnamefont {B.~H.}\ \bibnamefont
  {Kolner}},\ }\href {\doibase 10.1063/1.99181} {\bibfield  {journal} {\bibinfo
   {journal} {Appl. Phys. Lett.}\ }\textbf {\bibinfo {volume} {52}},\ \bibinfo
  {pages} {1122} (\bibinfo {year} {1988})}\BibitemShut {NoStop}%
\bibitem [{\citenamefont {Grischkowsky}(1974)}]{grischkowsky_optical_1974}%
  \BibitemOpen
  \bibfield  {author} {\bibinfo {author} {\bibfnamefont {D.}~\bibnamefont
  {Grischkowsky}},\ }\href {\doibase 10.1063/1.1655312} {\bibfield  {journal}
  {\bibinfo  {journal} {Appl. Phys. Lett.}\ }\textbf {\bibinfo {volume} {25}},\
  \bibinfo {pages} {566} (\bibinfo {year} {1974})}\BibitemShut {NoStop}%
\bibitem [{\citenamefont {Karpi\'{n}ski}\ \emph {et~al.}(2017)\citenamefont
  {Karpi\'{n}ski}, \citenamefont {Jachura}, \citenamefont {Wright},\ and\
  \citenamefont {Smith}}]{karpinski_bandwidth_2017}%
  \BibitemOpen
  \bibfield  {author} {\bibinfo {author} {\bibfnamefont {M.}~\bibnamefont
  {Karpi\'{n}ski}}, \bibinfo {author} {\bibfnamefont {M.}~\bibnamefont
  {Jachura}}, \bibinfo {author} {\bibfnamefont {L.~J.}\ \bibnamefont {Wright}},
  \ and\ \bibinfo {author} {\bibfnamefont {B.~J.}\ \bibnamefont {Smith}},\
  }\href {\doibase 10.1038/nphoton.2016.228} {\bibfield  {journal} {\bibinfo
  {journal} {Nat. Photonics}\ }\textbf {\bibinfo {volume} {11}},\ \bibinfo
  {pages} {53} (\bibinfo {year} {2017})},\ \Eprint
  {http://arxiv.org/abs/1604.02459} {arXiv:1604.02459} \BibitemShut {NoStop}%
\bibitem [{\citenamefont {Hernandez}\ \emph {et~al.}(2013)\citenamefont
  {Hernandez}, \citenamefont {Bennett}, \citenamefont {Moran}, \citenamefont
  {Drobshoff}, \citenamefont {Chang}, \citenamefont {Langrock}, \citenamefont
  {Fejer},\ and\ \citenamefont {Ibsen}}]{hernandez_104_2013}%
  \BibitemOpen
  \bibfield  {author} {\bibinfo {author} {\bibfnamefont {V.~J.}\ \bibnamefont
  {Hernandez}}, \bibinfo {author} {\bibfnamefont {C.~V.}\ \bibnamefont
  {Bennett}}, \bibinfo {author} {\bibfnamefont {B.~D.}\ \bibnamefont {Moran}},
  \bibinfo {author} {\bibfnamefont {A.~D.}\ \bibnamefont {Drobshoff}}, \bibinfo
  {author} {\bibfnamefont {D.}~\bibnamefont {Chang}}, \bibinfo {author}
  {\bibfnamefont {C.}~\bibnamefont {Langrock}}, \bibinfo {author}
  {\bibfnamefont {M.~M.}\ \bibnamefont {Fejer}}, \ and\ \bibinfo {author}
  {\bibfnamefont {M.}~\bibnamefont {Ibsen}},\ }\href {\doibase
  10.1364/oe.21.000196} {\bibfield  {journal} {\bibinfo  {journal} {Opt.
  Express}\ }\textbf {\bibinfo {volume} {21}},\ \bibinfo {pages} {196}
  (\bibinfo {year} {2013})}\BibitemShut {NoStop}%
\bibitem [{\citenamefont {Bennett}\ and\ \citenamefont
  {Kolner}(2001)}]{bennett_aberrations_2001}%
  \BibitemOpen
  \bibfield  {author} {\bibinfo {author} {\bibfnamefont {C.~V.}\ \bibnamefont
  {Bennett}}\ and\ \bibinfo {author} {\bibfnamefont {B.~H.}\ \bibnamefont
  {Kolner}},\ }\href {\doibase 10.1109/3.892720} {\bibfield  {journal}
  {\bibinfo  {journal} {IEEE J. Quantum Electron.}\ }\textbf {\bibinfo {volume}
  {37}},\ \bibinfo {pages} {20} (\bibinfo {year} {2001})}\BibitemShut {NoStop}%
\bibitem [{\citenamefont {Bennett}\ \emph {et~al.}(1994)\citenamefont
  {Bennett}, \citenamefont {Scott},\ and\ \citenamefont
  {Kolner}}]{bennett_temporal_1994}%
  \BibitemOpen
  \bibfield  {author} {\bibinfo {author} {\bibfnamefont {C.~V.}\ \bibnamefont
  {Bennett}}, \bibinfo {author} {\bibfnamefont {R.~P.}\ \bibnamefont {Scott}},
  \ and\ \bibinfo {author} {\bibfnamefont {B.~H.}\ \bibnamefont {Kolner}},\
  }\href {\doibase 10.1063/1.112620} {\bibfield  {journal} {\bibinfo  {journal}
  {Appl. Phys. Lett.}\ }\textbf {\bibinfo {volume} {65}},\ \bibinfo {pages}
  {2513} (\bibinfo {year} {1994})}\BibitemShut {NoStop}%
\bibitem [{\citenamefont {Bennett}\ and\ \citenamefont
  {Kolner}(1999)}]{bennett_upconversion_1999}%
  \BibitemOpen
  \bibfield  {author} {\bibinfo {author} {\bibfnamefont {C.~V.}\ \bibnamefont
  {Bennett}}\ and\ \bibinfo {author} {\bibfnamefont {B.~H.}\ \bibnamefont
  {Kolner}},\ }\href {\doibase 10.1364/ol.24.000783} {\bibfield  {journal}
  {\bibinfo  {journal} {Opt. Lett.}\ }\textbf {\bibinfo {volume} {24}},\
  \bibinfo {pages} {783} (\bibinfo {year} {1999})}\BibitemShut {NoStop}%
\bibitem [{\citenamefont {Agrawal}\ \emph {et~al.}(1989)\citenamefont
  {Agrawal}, \citenamefont {Baldeck},\ and\ \citenamefont
  {Alfano}}]{agrawal_temporal_1989}%
  \BibitemOpen
  \bibfield  {author} {\bibinfo {author} {\bibfnamefont {G.~P.}\ \bibnamefont
  {Agrawal}}, \bibinfo {author} {\bibfnamefont {P.~L.}\ \bibnamefont
  {Baldeck}}, \ and\ \bibinfo {author} {\bibfnamefont {R.~R.}\ \bibnamefont
  {Alfano}},\ }\href {\doibase 10.1103/PhysRevA.40.5063} {\bibfield  {journal}
  {\bibinfo  {journal} {Phys. Rev. A}\ }\textbf {\bibinfo {volume} {40}},\
  \bibinfo {pages} {5063} (\bibinfo {year} {1989})}\BibitemShut {NoStop}%
\bibitem [{\citenamefont {Kuzucu}\ \emph {et~al.}(2009)\citenamefont {Kuzucu},
  \citenamefont {Okawachi}, \citenamefont {Salem}, \citenamefont {Foster},
  \citenamefont {Turner-Foster}, \citenamefont {Lipson},\ and\ \citenamefont
  {Gaeta}}]{kuzucu_spectral_2009}%
  \BibitemOpen
  \bibfield  {author} {\bibinfo {author} {\bibfnamefont {O.}~\bibnamefont
  {Kuzucu}}, \bibinfo {author} {\bibfnamefont {Y.}~\bibnamefont {Okawachi}},
  \bibinfo {author} {\bibfnamefont {R.}~\bibnamefont {Salem}}, \bibinfo
  {author} {\bibfnamefont {M.~A.}\ \bibnamefont {Foster}}, \bibinfo {author}
  {\bibfnamefont {A.~C.}\ \bibnamefont {Turner-Foster}}, \bibinfo {author}
  {\bibfnamefont {M.}~\bibnamefont {Lipson}}, \ and\ \bibinfo {author}
  {\bibfnamefont {A.~L.}\ \bibnamefont {Gaeta}},\ }\href {\doibase
  10.1364/oe.17.020605} {\bibfield  {journal} {\bibinfo  {journal} {Opt.
  Express}\ }\textbf {\bibinfo {volume} {17}},\ \bibinfo {pages} {20605}
  (\bibinfo {year} {2009})}\BibitemShut {NoStop}%
\bibitem [{\citenamefont {Okawachi}\ \emph {et~al.}(2009)\citenamefont
  {Okawachi}, \citenamefont {Salem}, \citenamefont {Foster}, \citenamefont
  {Turner-Foster}, \citenamefont {Lipson},\ and\ \citenamefont
  {Gaeta}}]{okawachi_high-resolution_2009}%
  \BibitemOpen
  \bibfield  {author} {\bibinfo {author} {\bibfnamefont {Y.}~\bibnamefont
  {Okawachi}}, \bibinfo {author} {\bibfnamefont {R.}~\bibnamefont {Salem}},
  \bibinfo {author} {\bibfnamefont {M.~A.}\ \bibnamefont {Foster}}, \bibinfo
  {author} {\bibfnamefont {A.~C.}\ \bibnamefont {Turner-Foster}}, \bibinfo
  {author} {\bibfnamefont {M.}~\bibnamefont {Lipson}}, \ and\ \bibinfo {author}
  {\bibfnamefont {A.~L.}\ \bibnamefont {Gaeta}},\ }\href {\doibase
  10.1364/oe.17.005691} {\bibfield  {journal} {\bibinfo  {journal} {Opt.
  Express}\ }\textbf {\bibinfo {volume} {17}},\ \bibinfo {pages} {5691}
  (\bibinfo {year} {2009})}\BibitemShut {NoStop}%
\bibitem [{\citenamefont {Foster}\ \emph {et~al.}(2008)\citenamefont {Foster},
  \citenamefont {Salem}, \citenamefont {Geraghty}, \citenamefont
  {Turner-Foster}, \citenamefont {Lipson},\ and\ \citenamefont
  {Gaeta}}]{foster_silicon-chip-based_2008}%
  \BibitemOpen
  \bibfield  {author} {\bibinfo {author} {\bibfnamefont {M.~A.}\ \bibnamefont
  {Foster}}, \bibinfo {author} {\bibfnamefont {R.}~\bibnamefont {Salem}},
  \bibinfo {author} {\bibfnamefont {D.~F.}\ \bibnamefont {Geraghty}}, \bibinfo
  {author} {\bibfnamefont {A.~C.}\ \bibnamefont {Turner-Foster}}, \bibinfo
  {author} {\bibfnamefont {M.}~\bibnamefont {Lipson}}, \ and\ \bibinfo {author}
  {\bibfnamefont {A.~L.}\ \bibnamefont {Gaeta}},\ }\href {\doibase
  10.1038/nature07430} {\bibfield  {journal} {\bibinfo  {journal} {Nature}\
  }\textbf {\bibinfo {volume} {456}},\ \bibinfo {pages} {81} (\bibinfo {year}
  {2008})}\BibitemShut {NoStop}%
\bibitem [{\citenamefont {Foster}\ \emph {et~al.}(2009)\citenamefont {Foster},
  \citenamefont {Salem}, \citenamefont {Okawachi}, \citenamefont
  {Turner-Foster}, \citenamefont {Lipson},\ and\ \citenamefont
  {Gaeta}}]{foster_ultrafast_2009}%
  \BibitemOpen
  \bibfield  {author} {\bibinfo {author} {\bibfnamefont {M.~A.}\ \bibnamefont
  {Foster}}, \bibinfo {author} {\bibfnamefont {R.}~\bibnamefont {Salem}},
  \bibinfo {author} {\bibfnamefont {Y.}~\bibnamefont {Okawachi}}, \bibinfo
  {author} {\bibfnamefont {A.~C.}\ \bibnamefont {Turner-Foster}}, \bibinfo
  {author} {\bibfnamefont {M.}~\bibnamefont {Lipson}}, \ and\ \bibinfo {author}
  {\bibfnamefont {A.~L.}\ \bibnamefont {Gaeta}},\ }\href {\doibase
  10.1038/nphoton.2009.169} {\bibfield  {journal} {\bibinfo  {journal} {Nat.
  Photonics}\ }\textbf {\bibinfo {volume} {3}},\ \bibinfo {pages} {581}
  (\bibinfo {year} {2009})}\BibitemShut {NoStop}%
\bibitem [{\citenamefont {Li}\ \emph {et~al.}(2015)\citenamefont {Li},
  \citenamefont {Fern{\'{a}}ndez-Ruiz}, \citenamefont {Lou},\ and\
  \citenamefont {Aza{\~{n}}a}}]{li_high-contrast_2015}%
  \BibitemOpen
  \bibfield  {author} {\bibinfo {author} {\bibfnamefont {B.}~\bibnamefont
  {Li}}, \bibinfo {author} {\bibfnamefont {M.~R.}\ \bibnamefont
  {Fern{\'{a}}ndez-Ruiz}}, \bibinfo {author} {\bibfnamefont {S.}~\bibnamefont
  {Lou}}, \ and\ \bibinfo {author} {\bibfnamefont {J.}~\bibnamefont
  {Aza{\~{n}}a}},\ }\href {\doibase 10.1364/oe.23.006833} {\bibfield  {journal}
  {\bibinfo  {journal} {Opt. Express}\ }\textbf {\bibinfo {volume} {23}},\
  \bibinfo {pages} {6833} (\bibinfo {year} {2015})}\BibitemShut {NoStop}%
\bibitem [{\citenamefont {Donohue}\ \emph {et~al.}(2016)\citenamefont
  {Donohue}, \citenamefont {Mastrovich},\ and\ \citenamefont
  {Resch}}]{donohue_spectrally_2016}%
  \BibitemOpen
  \bibfield  {author} {\bibinfo {author} {\bibfnamefont {J.~M.}\ \bibnamefont
  {Donohue}}, \bibinfo {author} {\bibfnamefont {M.}~\bibnamefont {Mastrovich}},
  \ and\ \bibinfo {author} {\bibfnamefont {K.~J.}\ \bibnamefont {Resch}},\
  }\href {\doibase 10.1103/PhysRevLett.117.243602} {\bibfield  {journal}
  {\bibinfo  {journal} {Phys. Rev. Lett.}\ }\textbf {\bibinfo {volume} {117}},\
  \bibinfo {pages} {243602} (\bibinfo {year} {2016})},\ \Eprint
  {http://arxiv.org/abs/1604.03588} {arXiv:1604.03588} \BibitemShut {NoStop}%
\bibitem [{\citenamefont {Lu}\ \emph {et~al.}(2018)\citenamefont {Lu},
  \citenamefont {Lukens}, \citenamefont {Peters}, \citenamefont {Odele},
  \citenamefont {Leaird}, \citenamefont {Weiner},\ and\ \citenamefont
  {Lougovski}}]{Lu2018}%
  \BibitemOpen
  \bibfield  {author} {\bibinfo {author} {\bibfnamefont {H.~H.}\ \bibnamefont
  {Lu}}, \bibinfo {author} {\bibfnamefont {J.~M.}\ \bibnamefont {Lukens}},
  \bibinfo {author} {\bibfnamefont {N.~A.}\ \bibnamefont {Peters}}, \bibinfo
  {author} {\bibfnamefont {O.~D.}\ \bibnamefont {Odele}}, \bibinfo {author}
  {\bibfnamefont {D.~E.}\ \bibnamefont {Leaird}}, \bibinfo {author}
  {\bibfnamefont {A.~M.}\ \bibnamefont {Weiner}}, \ and\ \bibinfo {author}
  {\bibfnamefont {P.}~\bibnamefont {Lougovski}},\ }\href {\doibase
  10.1103/PhysRevLett.120.030502} {\bibfield  {journal} {\bibinfo  {journal}
  {Phys. Rev. Lett.}\ }\textbf {\bibinfo {volume} {120}},\ \bibinfo {pages}
  {030502} (\bibinfo {year} {2018})},\ \Eprint
  {http://arxiv.org/abs/1712.03992} {arXiv:1712.03992} \BibitemShut {NoStop}%
\bibitem [{\citenamefont {Denis}\ \emph {et~al.}(2017)\citenamefont {Denis},
  \citenamefont {Moreau}, \citenamefont {Devaux},\ and\ \citenamefont
  {Lantz}}]{denis_temporal_2017}%
  \BibitemOpen
  \bibfield  {author} {\bibinfo {author} {\bibfnamefont {S.}~\bibnamefont
  {Denis}}, \bibinfo {author} {\bibfnamefont {P.~A.}\ \bibnamefont {Moreau}},
  \bibinfo {author} {\bibfnamefont {F.}~\bibnamefont {Devaux}}, \ and\ \bibinfo
  {author} {\bibfnamefont {E.}~\bibnamefont {Lantz}},\ }\href {\doibase
  10.1088/2040-8986/aa587b} {\bibfield  {journal} {\bibinfo  {journal} {J.
  Opt.}\ }\textbf {\bibinfo {volume} {19}},\ \bibinfo {pages} {34002} (\bibinfo
  {year} {2017})},\ \Eprint {http://arxiv.org/abs/1612.05723}
  {arXiv:1612.05723} \BibitemShut {NoStop}%
\bibitem [{\citenamefont {Dong}\ \emph {et~al.}(2016)\citenamefont {Dong},
  \citenamefont {Zhang}, \citenamefont {Huang},\ and\ \citenamefont
  {Peng}}]{dong_long-distance_2016}%
  \BibitemOpen
  \bibfield  {author} {\bibinfo {author} {\bibfnamefont {S.}~\bibnamefont
  {Dong}}, \bibinfo {author} {\bibfnamefont {W.}~\bibnamefont {Zhang}},
  \bibinfo {author} {\bibfnamefont {Y.}~\bibnamefont {Huang}}, \ and\ \bibinfo
  {author} {\bibfnamefont {J.}~\bibnamefont {Peng}},\ }\href {\doibase
  10.1038/srep26022} {\bibfield  {journal} {\bibinfo  {journal} {Sci. Rep.}\
  }\textbf {\bibinfo {volume} {6}},\ \bibinfo {pages} {26022} (\bibinfo {year}
  {2016})},\ \Eprint {http://arxiv.org/abs/1508.05248} {arXiv:1508.05248}
  \BibitemShut {NoStop}%
\bibitem [{\citenamefont {Ryczkowski}\ \emph {et~al.}(2017)\citenamefont
  {Ryczkowski}, \citenamefont {Barbier}, \citenamefont {Friberg}, \citenamefont
  {Dudley},\ and\ \citenamefont {Genty}}]{ryczkowski_magnified_2017}%
  \BibitemOpen
  \bibfield  {author} {\bibinfo {author} {\bibfnamefont {P.}~\bibnamefont
  {Ryczkowski}}, \bibinfo {author} {\bibfnamefont {M.}~\bibnamefont {Barbier}},
  \bibinfo {author} {\bibfnamefont {A.~T.}\ \bibnamefont {Friberg}}, \bibinfo
  {author} {\bibfnamefont {J.~M.}\ \bibnamefont {Dudley}}, \ and\ \bibinfo
  {author} {\bibfnamefont {G.}~\bibnamefont {Genty}},\ }\href {\doibase
  10.1063/1.4977534} {\bibfield  {journal} {\bibinfo  {journal} {APL
  Photonics}\ }\textbf {\bibinfo {volume} {2}},\ \bibinfo {pages} {46102}
  (\bibinfo {year} {2017})},\ \Eprint {http://arxiv.org/abs/1701.00163}
  {arXiv:1701.00163} \BibitemShut {NoStop}%
\bibitem [{\citenamefont {Wu}\ \emph {et~al.}(2019)\citenamefont {Wu},
  \citenamefont {Ryczkowski}, \citenamefont {Friberg}, \citenamefont {Dudley},\
  and\ \citenamefont {Genty}}]{wu_temporal_2019}%
  \BibitemOpen
  \bibfield  {author} {\bibinfo {author} {\bibfnamefont {H.}~\bibnamefont
  {Wu}}, \bibinfo {author} {\bibfnamefont {P.}~\bibnamefont {Ryczkowski}},
  \bibinfo {author} {\bibfnamefont {A.~T.}\ \bibnamefont {Friberg}}, \bibinfo
  {author} {\bibfnamefont {J.~M.}\ \bibnamefont {Dudley}}, \ and\ \bibinfo
  {author} {\bibfnamefont {G.}~\bibnamefont {Genty}},\ }\href {\doibase
  10.1364/optica.6.000902} {\bibfield  {journal} {\bibinfo  {journal} {Optica}\
  }\textbf {\bibinfo {volume} {6}},\ \bibinfo {pages} {902} (\bibinfo {year}
  {2019})}\BibitemShut {NoStop}%
\bibitem [{\citenamefont {Allgaier}\ \emph {et~al.}(2017)\citenamefont
  {Allgaier}, \citenamefont {Ansari}, \citenamefont {Sansoni}, \citenamefont
  {Eigner}, \citenamefont {Quiring}, \citenamefont {Ricken}, \citenamefont
  {Harder}, \citenamefont {Brecht},\ and\ \citenamefont
  {Silberhorn}}]{allgaier_highly_2017}%
  \BibitemOpen
  \bibfield  {author} {\bibinfo {author} {\bibfnamefont {M.}~\bibnamefont
  {Allgaier}}, \bibinfo {author} {\bibfnamefont {V.}~\bibnamefont {Ansari}},
  \bibinfo {author} {\bibfnamefont {L.}~\bibnamefont {Sansoni}}, \bibinfo
  {author} {\bibfnamefont {C.}~\bibnamefont {Eigner}}, \bibinfo {author}
  {\bibfnamefont {V.}~\bibnamefont {Quiring}}, \bibinfo {author} {\bibfnamefont
  {R.}~\bibnamefont {Ricken}}, \bibinfo {author} {\bibfnamefont
  {G.}~\bibnamefont {Harder}}, \bibinfo {author} {\bibfnamefont
  {B.}~\bibnamefont {Brecht}}, \ and\ \bibinfo {author} {\bibfnamefont
  {C.}~\bibnamefont {Silberhorn}},\ }\href {\doibase 10.1038/ncomms14288}
  {\bibfield  {journal} {\bibinfo  {journal} {Nat. Commun.}\ }\textbf {\bibinfo
  {volume} {8}},\ \bibinfo {pages} {14288} (\bibinfo {year}
  {2017})}\BibitemShut {NoStop}%
\bibitem [{\citenamefont {Zhao}\ \emph {et~al.}(2014)\citenamefont {Zhao},
  \citenamefont {Guo}, \citenamefont {Liu}, \citenamefont {Sun}, \citenamefont
  {Loy},\ and\ \citenamefont {Du}}]{Zhao2014}%
  \BibitemOpen
  \bibfield  {author} {\bibinfo {author} {\bibfnamefont {L.}~\bibnamefont
  {Zhao}}, \bibinfo {author} {\bibfnamefont {X.}~\bibnamefont {Guo}}, \bibinfo
  {author} {\bibfnamefont {C.}~\bibnamefont {Liu}}, \bibinfo {author}
  {\bibfnamefont {Y.}~\bibnamefont {Sun}}, \bibinfo {author} {\bibfnamefont
  {M.~M.~T.}\ \bibnamefont {Loy}}, \ and\ \bibinfo {author} {\bibfnamefont
  {S.}~\bibnamefont {Du}},\ }\href {\doibase 10.1364/optica.1.000084}
  {\bibfield  {journal} {\bibinfo  {journal} {Optica}\ }\textbf {\bibinfo
  {volume} {1}},\ \bibinfo {pages} {84} (\bibinfo {year} {2014})}\BibitemShut
  {NoStop}%
\bibitem [{\citenamefont {Guo}\ \emph {et~al.}(2017)\citenamefont {Guo},
  \citenamefont {Mei},\ and\ \citenamefont {Du}}]{Guo2017}%
  \BibitemOpen
  \bibfield  {author} {\bibinfo {author} {\bibfnamefont {X.}~\bibnamefont
  {Guo}}, \bibinfo {author} {\bibfnamefont {Y.}~\bibnamefont {Mei}}, \ and\
  \bibinfo {author} {\bibfnamefont {S.}~\bibnamefont {Du}},\ }\href {\doibase
  10.1364/optica.4.000388} {\bibfield  {journal} {\bibinfo  {journal} {Optica}\
  }\textbf {\bibinfo {volume} {4}},\ \bibinfo {pages} {388} (\bibinfo {year}
  {2017})},\ \Eprint {http://arxiv.org/abs/1609.02282} {arXiv:1609.02282}
  \BibitemShut {NoStop}%
\bibitem [{\citenamefont {Farrera}\ \emph {et~al.}(2016)\citenamefont
  {Farrera}, \citenamefont {Heinze}, \citenamefont {Albrecht}, \citenamefont
  {Ho}, \citenamefont {Ch{\'{a}}vez}, \citenamefont {Teo}, \citenamefont
  {Sangouard},\ and\ \citenamefont {{De Riedmatten}}}]{Farrera2016}%
  \BibitemOpen
  \bibfield  {author} {\bibinfo {author} {\bibfnamefont {P.}~\bibnamefont
  {Farrera}}, \bibinfo {author} {\bibfnamefont {G.}~\bibnamefont {Heinze}},
  \bibinfo {author} {\bibfnamefont {B.}~\bibnamefont {Albrecht}}, \bibinfo
  {author} {\bibfnamefont {M.}~\bibnamefont {Ho}}, \bibinfo {author}
  {\bibfnamefont {M.}~\bibnamefont {Ch{\'{a}}vez}}, \bibinfo {author}
  {\bibfnamefont {C.}~\bibnamefont {Teo}}, \bibinfo {author} {\bibfnamefont
  {N.}~\bibnamefont {Sangouard}}, \ and\ \bibinfo {author} {\bibfnamefont
  {H.}~\bibnamefont {{De Riedmatten}}},\ }\href {\doibase 10.1038/ncomms13556}
  {\bibfield  {journal} {\bibinfo  {journal} {Nature Communications}\ }\textbf
  {\bibinfo {volume} {7}},\ \bibinfo {pages} {13556} (\bibinfo {year}
  {2016})},\ \Eprint {http://arxiv.org/abs/1601.07142} {arXiv:1601.07142}
  \BibitemShut {NoStop}%
\bibitem [{\citenamefont {Stute}\ \emph {et~al.}(2012)\citenamefont {Stute},
  \citenamefont {Casabone}, \citenamefont {Schindler}, \citenamefont {Monz},
  \citenamefont {Schmidt}, \citenamefont {Brandst{\"{a}}tter}, \citenamefont
  {Northup},\ and\ \citenamefont {Blatt}}]{Stute2012}%
  \BibitemOpen
  \bibfield  {author} {\bibinfo {author} {\bibfnamefont {A.}~\bibnamefont
  {Stute}}, \bibinfo {author} {\bibfnamefont {B.}~\bibnamefont {Casabone}},
  \bibinfo {author} {\bibfnamefont {P.}~\bibnamefont {Schindler}}, \bibinfo
  {author} {\bibfnamefont {T.}~\bibnamefont {Monz}}, \bibinfo {author}
  {\bibfnamefont {P.~O.}\ \bibnamefont {Schmidt}}, \bibinfo {author}
  {\bibfnamefont {B.}~\bibnamefont {Brandst{\"{a}}tter}}, \bibinfo {author}
  {\bibfnamefont {T.~E.}\ \bibnamefont {Northup}}, \ and\ \bibinfo {author}
  {\bibfnamefont {R.}~\bibnamefont {Blatt}},\ }\href {\doibase
  10.1038/nature11120} {\bibfield  {journal} {\bibinfo  {journal} {Nature}\
  }\textbf {\bibinfo {volume} {485}},\ \bibinfo {pages} {482} (\bibinfo {year}
  {2012})}\BibitemShut {NoStop}%
\bibitem [{\citenamefont {Rambach}\ \emph {et~al.}(2016)\citenamefont
  {Rambach}, \citenamefont {Nikolova}, \citenamefont {Weinhold},\ and\
  \citenamefont {White}}]{Rambach2016}%
  \BibitemOpen
  \bibfield  {author} {\bibinfo {author} {\bibfnamefont {M.}~\bibnamefont
  {Rambach}}, \bibinfo {author} {\bibfnamefont {A.}~\bibnamefont {Nikolova}},
  \bibinfo {author} {\bibfnamefont {T.~J.}\ \bibnamefont {Weinhold}}, \ and\
  \bibinfo {author} {\bibfnamefont {A.~G.}\ \bibnamefont {White}},\ }\href
  {\doibase 10.1063/1.4966915} {\bibfield  {journal} {\bibinfo  {journal} {APL
  Photonics}\ }\textbf {\bibinfo {volume} {1}},\ \bibinfo {pages} {096101}
  (\bibinfo {year} {2016})},\ \Eprint {http://arxiv.org/abs/1601.06173}
  {arXiv:1601.06173} \BibitemShut {NoStop}%
\bibitem [{\citenamefont {Hong}\ \emph {et~al.}(2017)\citenamefont {Hong},
  \citenamefont {Riedinger}, \citenamefont {Marinkovi{\'{c}}}, \citenamefont
  {Wallucks}, \citenamefont {Hofer}, \citenamefont {Norte}, \citenamefont
  {Aspelmeyer},\ and\ \citenamefont {Gr{\"{o}}blacher}}]{Hong2017}%
  \BibitemOpen
  \bibfield  {author} {\bibinfo {author} {\bibfnamefont {S.}~\bibnamefont
  {Hong}}, \bibinfo {author} {\bibfnamefont {R.}~\bibnamefont {Riedinger}},
  \bibinfo {author} {\bibfnamefont {I.}~\bibnamefont {Marinkovi{\'{c}}}},
  \bibinfo {author} {\bibfnamefont {A.}~\bibnamefont {Wallucks}}, \bibinfo
  {author} {\bibfnamefont {S.~G.}\ \bibnamefont {Hofer}}, \bibinfo {author}
  {\bibfnamefont {R.~A.}\ \bibnamefont {Norte}}, \bibinfo {author}
  {\bibfnamefont {M.}~\bibnamefont {Aspelmeyer}}, \ and\ \bibinfo {author}
  {\bibfnamefont {S.}~\bibnamefont {Gr{\"{o}}blacher}},\ }\href {\doibase
  10.1126/science.aan7939} {\bibfield  {journal} {\bibinfo  {journal}
  {Science}\ }\textbf {\bibinfo {volume} {358}},\ \bibinfo {pages} {203}
  (\bibinfo {year} {2017})},\ \Eprint {http://arxiv.org/abs/1706.03777}
  {arXiv:1706.03777} \BibitemShut {NoStop}%
\bibitem [{\citenamefont {Hill}\ \emph {et~al.}(2012)\citenamefont {Hill},
  \citenamefont {Safavi-Naeini}, \citenamefont {Chan},\ and\ \citenamefont
  {Painter}}]{Hill2012}%
  \BibitemOpen
  \bibfield  {author} {\bibinfo {author} {\bibfnamefont {J.~T.}\ \bibnamefont
  {Hill}}, \bibinfo {author} {\bibfnamefont {A.~H.}\ \bibnamefont
  {Safavi-Naeini}}, \bibinfo {author} {\bibfnamefont {J.}~\bibnamefont {Chan}},
  \ and\ \bibinfo {author} {\bibfnamefont {O.}~\bibnamefont {Painter}},\ }\href
  {\doibase 10.1038/ncomms2201} {\bibfield  {journal} {\bibinfo  {journal}
  {Nature Communications}\ }\textbf {\bibinfo {volume} {3}},\ \bibinfo {pages}
  {1196} (\bibinfo {year} {2012})},\ \Eprint {http://arxiv.org/abs/1206.0704}
  {arXiv:1206.0704} \BibitemShut {NoStop}%
\bibitem [{\citenamefont {Mei}\ \emph {et~al.}(2020)\citenamefont {Mei},
  \citenamefont {Zhou}, \citenamefont {Zhang}, \citenamefont {Li},
  \citenamefont {Liao}, \citenamefont {Yan}, \citenamefont {Zhu},\ and\
  \citenamefont {Du}}]{Mei2019}%
  \BibitemOpen
  \bibfield  {author} {\bibinfo {author} {\bibfnamefont {Y.}~\bibnamefont
  {Mei}}, \bibinfo {author} {\bibfnamefont {Y.}~\bibnamefont {Zhou}}, \bibinfo
  {author} {\bibfnamefont {S.}~\bibnamefont {Zhang}}, \bibinfo {author}
  {\bibfnamefont {J.}~\bibnamefont {Li}}, \bibinfo {author} {\bibfnamefont
  {K.}~\bibnamefont {Liao}}, \bibinfo {author} {\bibfnamefont {H.}~\bibnamefont
  {Yan}}, \bibinfo {author} {\bibfnamefont {S.-L.}\ \bibnamefont {Zhu}}, \ and\
  \bibinfo {author} {\bibfnamefont {S.}~\bibnamefont {Du}},\ }\href {\doibase
  10.1103/PhysRevLett.124.010509} {\bibfield  {journal} {\bibinfo  {journal}
  {Phys. Rev. Lett.}\ }\textbf {\bibinfo {volume} {124}},\ \bibinfo {pages}
  {010509} (\bibinfo {year} {2020})}\BibitemShut {NoStop}%
\bibitem [{\citenamefont {Hosseini}\ \emph {et~al.}(2009)\citenamefont
  {Hosseini}, \citenamefont {Sparkes}, \citenamefont {H{\'{e}}tet},
  \citenamefont {Longdell}, \citenamefont {Lam},\ and\ \citenamefont
  {Buchler}}]{Hosseini2009}%
  \BibitemOpen
  \bibfield  {author} {\bibinfo {author} {\bibfnamefont {M.}~\bibnamefont
  {Hosseini}}, \bibinfo {author} {\bibfnamefont {B.~M.}\ \bibnamefont
  {Sparkes}}, \bibinfo {author} {\bibfnamefont {G.}~\bibnamefont
  {H{\'{e}}tet}}, \bibinfo {author} {\bibfnamefont {J.~J.}\ \bibnamefont
  {Longdell}}, \bibinfo {author} {\bibfnamefont {P.~K.}\ \bibnamefont {Lam}}, \
  and\ \bibinfo {author} {\bibfnamefont {B.~C.}\ \bibnamefont {Buchler}},\
  }\href {http://www.nature.com/articles/nature08325} {\bibfield  {journal}
  {\bibinfo  {journal} {Nature}\ }\textbf {\bibinfo {volume} {461}},\ \bibinfo
  {pages} {241} (\bibinfo {year} {2009})}\BibitemShut {NoStop}%
\bibitem [{\citenamefont {Kauffman}\ \emph {et~al.}(1994)\citenamefont
  {Kauffman}, \citenamefont {Banyai}, \citenamefont {Godil},\ and\
  \citenamefont {Bloom}}]{Kauffman1994}%
  \BibitemOpen
  \bibfield  {author} {\bibinfo {author} {\bibfnamefont {M.~T.}\ \bibnamefont
  {Kauffman}}, \bibinfo {author} {\bibfnamefont {W.~C.}\ \bibnamefont
  {Banyai}}, \bibinfo {author} {\bibfnamefont {A.~A.}\ \bibnamefont {Godil}}, \
  and\ \bibinfo {author} {\bibfnamefont {D.~M.}\ \bibnamefont {Bloom}},\ }\href
  {\doibase 10.1063/1.111177} {\bibfield  {journal} {\bibinfo  {journal} {Appl.
  Phys. Lett.}\ }\textbf {\bibinfo {volume} {64}},\ \bibinfo {pages} {270}
  (\bibinfo {year} {1994})}\BibitemShut {NoStop}%
\bibitem [{\citenamefont {Aza{\~{n}}a}\ \emph {et~al.}(2004)\citenamefont
  {Aza{\~{n}}a}, \citenamefont {Berger}, \citenamefont {Levit},\ and\
  \citenamefont {Fischer}}]{Azana2004}%
  \BibitemOpen
  \bibfield  {author} {\bibinfo {author} {\bibfnamefont {J.}~\bibnamefont
  {Aza{\~{n}}a}}, \bibinfo {author} {\bibfnamefont {N.~K.}\ \bibnamefont
  {Berger}}, \bibinfo {author} {\bibfnamefont {B.}~\bibnamefont {Levit}}, \
  and\ \bibinfo {author} {\bibfnamefont {B.}~\bibnamefont {Fischer}},\ }\href
  {\doibase 10.1364/AO.43.000483} {\bibfield  {journal} {\bibinfo  {journal}
  {Appl. Opt.}\ }\textbf {\bibinfo {volume} {43}},\ \bibinfo {pages} {483}
  (\bibinfo {year} {2004})}\BibitemShut {NoStop}%
\bibitem [{\citenamefont {Babashah}\ \emph {et~al.}(2019)\citenamefont
  {Babashah}, \citenamefont {Kavehvash}, \citenamefont {Khavasi},\ and\
  \citenamefont {Koohi}}]{Babashah2019}%
  \BibitemOpen
  \bibfield  {author} {\bibinfo {author} {\bibfnamefont {H.}~\bibnamefont
  {Babashah}}, \bibinfo {author} {\bibfnamefont {Z.}~\bibnamefont {Kavehvash}},
  \bibinfo {author} {\bibfnamefont {A.}~\bibnamefont {Khavasi}}, \ and\
  \bibinfo {author} {\bibfnamefont {S.}~\bibnamefont {Koohi}},\ }\href
  {\doibase 10.1016/j.optlastec.2018.09.027} {\bibfield  {journal} {\bibinfo
  {journal} {Opt. Laser Technol.}\ }\textbf {\bibinfo {volume} {111}},\
  \bibinfo {pages} {66} (\bibinfo {year} {2019})},\ \Eprint
  {http://arxiv.org/abs/1712.06482} {arXiv:1712.06482} \BibitemShut {NoStop}%
\bibitem [{\citenamefont {Salem}\ \emph {et~al.}(2009)\citenamefont {Salem},
  \citenamefont {Foster}, \citenamefont {Turner-Foster}, \citenamefont
  {Geraghty}, \citenamefont {Lipson},\ and\ \citenamefont {Gaeta}}]{Salem2009}%
  \BibitemOpen
  \bibfield  {author} {\bibinfo {author} {\bibfnamefont {R.}~\bibnamefont
  {Salem}}, \bibinfo {author} {\bibfnamefont {M.~A.}\ \bibnamefont {Foster}},
  \bibinfo {author} {\bibfnamefont {A.~C.}\ \bibnamefont {Turner-Foster}},
  \bibinfo {author} {\bibfnamefont {D.~F.}\ \bibnamefont {Geraghty}}, \bibinfo
  {author} {\bibfnamefont {M.}~\bibnamefont {Lipson}}, \ and\ \bibinfo {author}
  {\bibfnamefont {A.~L.}\ \bibnamefont {Gaeta}},\ }\href {\doibase
  10.1364/oe.17.004324} {\bibfield  {journal} {\bibinfo  {journal} {Opt.
  Express}\ }\textbf {\bibinfo {volume} {17}},\ \bibinfo {pages} {4324}
  (\bibinfo {year} {2009})}\BibitemShut {NoStop}%
\bibitem [{\citenamefont {Suret}\ \emph {et~al.}(2016)\citenamefont {Suret},
  \citenamefont {{El Koussaifi}}, \citenamefont {Tikan}, \citenamefont {Evain},
  \citenamefont {Randoux}, \citenamefont {Szwaj},\ and\ \citenamefont
  {Bielawski}}]{Suret2016}%
  \BibitemOpen
  \bibfield  {author} {\bibinfo {author} {\bibfnamefont {P.}~\bibnamefont
  {Suret}}, \bibinfo {author} {\bibfnamefont {R.}~\bibnamefont {{El
  Koussaifi}}}, \bibinfo {author} {\bibfnamefont {A.}~\bibnamefont {Tikan}},
  \bibinfo {author} {\bibfnamefont {C.}~\bibnamefont {Evain}}, \bibinfo
  {author} {\bibfnamefont {S.}~\bibnamefont {Randoux}}, \bibinfo {author}
  {\bibfnamefont {C.}~\bibnamefont {Szwaj}}, \ and\ \bibinfo {author}
  {\bibfnamefont {S.}~\bibnamefont {Bielawski}},\ }\href {\doibase
  10.1038/ncomms13136} {\bibfield  {journal} {\bibinfo  {journal} {Nat.
  Commun.}\ }\textbf {\bibinfo {volume} {7}},\ \bibinfo {pages} {13136}
  (\bibinfo {year} {2016})},\ \Eprint {http://arxiv.org/abs/1603.01477}
  {arXiv:1603.01477} \BibitemShut {NoStop}%
\bibitem [{\citenamefont {Mouradian}\ \emph {et~al.}(2000)\citenamefont
  {Mouradian}, \citenamefont {Louradour}, \citenamefont {Messager},
  \citenamefont {Barth{\'{e}}l{\'{e}}my},\ and\ \citenamefont
  {Froehly}}]{Mouradian2000}%
  \BibitemOpen
  \bibfield  {author} {\bibinfo {author} {\bibfnamefont {L.~K.}\ \bibnamefont
  {Mouradian}}, \bibinfo {author} {\bibfnamefont {F.}~\bibnamefont
  {Louradour}}, \bibinfo {author} {\bibfnamefont {V.}~\bibnamefont {Messager}},
  \bibinfo {author} {\bibfnamefont {A.}~\bibnamefont {Barth{\'{e}}l{\'{e}}my}},
  \ and\ \bibinfo {author} {\bibfnamefont {C.}~\bibnamefont {Froehly}},\ }\href
  {\doibase 10.1109/3.848351} {\bibfield  {journal} {\bibinfo  {journal} {IEEE
  J. Quantum Electron.}\ }\textbf {\bibinfo {volume} {36}},\ \bibinfo {pages}
  {795} (\bibinfo {year} {2000})}\BibitemShut {NoStop}%
\bibitem [{\citenamefont {Sparkes}\ \emph {et~al.}(2013)\citenamefont
  {Sparkes}, \citenamefont {Bernu}, \citenamefont {Hosseini}, \citenamefont
  {Geng}, \citenamefont {Glorieux}, \citenamefont {Altin}, \citenamefont {Lam},
  \citenamefont {Robins},\ and\ \citenamefont {Buchler}}]{Sparkes2013}%
  \BibitemOpen
  \bibfield  {author} {\bibinfo {author} {\bibfnamefont {B.~M.}\ \bibnamefont
  {Sparkes}}, \bibinfo {author} {\bibfnamefont {J.}~\bibnamefont {Bernu}},
  \bibinfo {author} {\bibfnamefont {M.}~\bibnamefont {Hosseini}}, \bibinfo
  {author} {\bibfnamefont {J.}~\bibnamefont {Geng}}, \bibinfo {author}
  {\bibfnamefont {Q.}~\bibnamefont {Glorieux}}, \bibinfo {author}
  {\bibfnamefont {P.~A.}\ \bibnamefont {Altin}}, \bibinfo {author}
  {\bibfnamefont {P.~K.}\ \bibnamefont {Lam}}, \bibinfo {author} {\bibfnamefont
  {N.~P.}\ \bibnamefont {Robins}}, \ and\ \bibinfo {author} {\bibfnamefont
  {B.~C.}\ \bibnamefont {Buchler}},\ }\href {\doibase
  10.1088/1367-2630/15/8/085027} {\bibfield  {journal} {\bibinfo  {journal}
  {New J. Phys.}\ }\textbf {\bibinfo {volume} {15}},\ \bibinfo {pages} {085027}
  (\bibinfo {year} {2013})},\ \Eprint {http://arxiv.org/abs/1211.7171}
  {arXiv:1211.7171} \BibitemShut {NoStop}%
\bibitem [{\citenamefont {Leszczy{\'{n}}ski}\ \emph {et~al.}(2018)\citenamefont
  {Leszczy{\'{n}}ski}, \citenamefont {Mazelanik}, \citenamefont {Lipka},
  \citenamefont {Parniak}, \citenamefont {Dąbrowski},\ and\ \citenamefont
  {Wasilewski}}]{Leszczynski2018}%
  \BibitemOpen
  \bibfield  {author} {\bibinfo {author} {\bibfnamefont {A.}~\bibnamefont
  {Leszczy{\'{n}}ski}}, \bibinfo {author} {\bibfnamefont {M.}~\bibnamefont
  {Mazelanik}}, \bibinfo {author} {\bibfnamefont {M.}~\bibnamefont {Lipka}},
  \bibinfo {author} {\bibfnamefont {M.}~\bibnamefont {Parniak}}, \bibinfo
  {author} {\bibfnamefont {M.}~\bibnamefont {Dąbrowski}}, \ and\ \bibinfo
  {author} {\bibfnamefont {W.}~\bibnamefont {Wasilewski}},\ }\href {\doibase
  10.1364/ol.43.001147} {\bibfield  {journal} {\bibinfo  {journal} {Opt.
  Lett.}\ }\textbf {\bibinfo {volume} {43}},\ \bibinfo {pages} {1147} (\bibinfo
  {year} {2018})},\ \Eprint {http://arxiv.org/abs/1712.07747}
  {arXiv:1712.07747} \BibitemShut {NoStop}%
\bibitem [{\citenamefont {Parniak}\ \emph {et~al.}(2019)\citenamefont
  {Parniak}, \citenamefont {Mazelanik}, \citenamefont {Leszczy{\'{n}}ski},
  \citenamefont {Lipka}, \citenamefont {Dąbrowski},\ and\ \citenamefont
  {Wasilewski}}]{Parniak2019}%
  \BibitemOpen
  \bibfield  {author} {\bibinfo {author} {\bibfnamefont {M.}~\bibnamefont
  {Parniak}}, \bibinfo {author} {\bibfnamefont {M.}~\bibnamefont {Mazelanik}},
  \bibinfo {author} {\bibfnamefont {A.}~\bibnamefont {Leszczy{\'{n}}ski}},
  \bibinfo {author} {\bibfnamefont {M.}~\bibnamefont {Lipka}}, \bibinfo
  {author} {\bibfnamefont {M.}~\bibnamefont {Dąbrowski}}, \ and\ \bibinfo
  {author} {\bibfnamefont {W.}~\bibnamefont {Wasilewski}},\ }\href {\doibase
  10.1103/PhysRevLett.122.063604} {\bibfield  {journal} {\bibinfo  {journal}
  {Phys. Rev. Lett.}\ }\textbf {\bibinfo {volume} {122}},\ \bibinfo {pages}
  {063604} (\bibinfo {year} {2019})},\ \Eprint
  {http://arxiv.org/abs/1804.05854} {arXiv:1804.05854} \BibitemShut {NoStop}%
\bibitem [{\citenamefont {Lipka}\ \emph {et~al.}(2019)\citenamefont {Lipka},
  \citenamefont {Leszczy{\'{n}}ski}, \citenamefont {Mazelanik}, \citenamefont
  {Parniak},\ and\ \citenamefont {Wasilewski}}]{Lipka2019}%
  \BibitemOpen
  \bibfield  {author} {\bibinfo {author} {\bibfnamefont {M.}~\bibnamefont
  {Lipka}}, \bibinfo {author} {\bibfnamefont {A.}~\bibnamefont
  {Leszczy{\'{n}}ski}}, \bibinfo {author} {\bibfnamefont {M.}~\bibnamefont
  {Mazelanik}}, \bibinfo {author} {\bibfnamefont {M.}~\bibnamefont {Parniak}},
  \ and\ \bibinfo {author} {\bibfnamefont {W.}~\bibnamefont {Wasilewski}},\
  }\href {\doibase 10.1103/PhysRevApplied.11.034049} {\bibfield  {journal}
  {\bibinfo  {journal} {Phys. Rev. Appl.}\ }\textbf {\bibinfo {volume} {11}},\
  \bibinfo {pages} {034049} (\bibinfo {year} {2019})}\BibitemShut {NoStop}%
\bibitem [{\citenamefont {Hedges}\ \emph {et~al.}(2010)\citenamefont {Hedges},
  \citenamefont {Longdell}, \citenamefont {Li},\ and\ \citenamefont
  {Sellars}}]{Hedges2010}%
  \BibitemOpen
  \bibfield  {author} {\bibinfo {author} {\bibfnamefont {M.~P.}\ \bibnamefont
  {Hedges}}, \bibinfo {author} {\bibfnamefont {J.~J.}\ \bibnamefont
  {Longdell}}, \bibinfo {author} {\bibfnamefont {Y.}~\bibnamefont {Li}}, \ and\
  \bibinfo {author} {\bibfnamefont {M.~J.}\ \bibnamefont {Sellars}},\ }\href
  {\doibase 10.1038/nature09081} {\bibfield  {journal} {\bibinfo  {journal}
  {Nature}\ }\textbf {\bibinfo {volume} {465}},\ \bibinfo {pages} {1052}
  (\bibinfo {year} {2010})}\BibitemShut {NoStop}%
\bibitem [{\citenamefont {Jeong}\ \emph {et~al.}(2019)\citenamefont {Jeong},
  \citenamefont {Du},\ and\ \citenamefont {Kim}}]{Jeong2019}%
  \BibitemOpen
  \bibfield  {author} {\bibinfo {author} {\bibfnamefont {H.}~\bibnamefont
  {Jeong}}, \bibinfo {author} {\bibfnamefont {S.}~\bibnamefont {Du}}, \ and\
  \bibinfo {author} {\bibfnamefont {N.~Y.}\ \bibnamefont {Kim}},\ }\href
  {\doibase 10.1364/josab.36.000646} {\bibfield  {journal} {\bibinfo  {journal}
  {Journal of the Optical Society of America B}\ }\textbf {\bibinfo {volume}
  {36}},\ \bibinfo {pages} {646} (\bibinfo {year} {2019})}\BibitemShut
  {NoStop}%
\bibitem [{\citenamefont {Saglamyurek}\ \emph {et~al.}(2018)\citenamefont
  {Saglamyurek}, \citenamefont {Hrushevskyi}, \citenamefont {Rastogi},
  \citenamefont {Heshami},\ and\ \citenamefont {LeBlanc}}]{Saglamyurek2018}%
  \BibitemOpen
  \bibfield  {author} {\bibinfo {author} {\bibfnamefont {E.}~\bibnamefont
  {Saglamyurek}}, \bibinfo {author} {\bibfnamefont {T.}~\bibnamefont
  {Hrushevskyi}}, \bibinfo {author} {\bibfnamefont {A.}~\bibnamefont
  {Rastogi}}, \bibinfo {author} {\bibfnamefont {K.}~\bibnamefont {Heshami}}, \
  and\ \bibinfo {author} {\bibfnamefont {L.~J.}\ \bibnamefont {LeBlanc}},\
  }\href {\doibase 10.1038/s41566-018-0279-0} {\bibfield  {journal} {\bibinfo
  {journal} {Nature Photonics}\ }\textbf {\bibinfo {volume} {12}},\ \bibinfo
  {pages} {774} (\bibinfo {year} {2018})},\ \Eprint
  {http://arxiv.org/abs/1710.08902} {arXiv:1710.08902} \BibitemShut {NoStop}%
\end{thebibliography}%
\clearpage

\renewcommand{\thefigure}{S\arabic{figure}}
\renewcommand{\theequation}{S\arabic{equation}}
\setcounter{figure}{0}
\setcounter{equation}{0}

\section*{Supplementary material}
\subsection*{Light-atoms interaction}

To describe interaction between light and atomic coherence we use
three level atom model with adiabatic elimination. The most comfortable
coordinate system runs in time with beam $t\to t+z/c$. We explicitly
make the control Rabi frequency $\Omega(t)$ time-dependent, as this
is directly controlled in the experiment. Notably, $\Omega(t)$ represents
the slowly-varying amplitude of this control field. Furthermore, we
write the equations in terms of demodulated zero-spatial-frequency
coherence $\check{\rho}_{hg}(z,t)=\rho_{hg}(z,t)e^{iK_{z0}z-i\Delta_{\mathrm{{HFS}}}t}$,
where $\rho_{hg}(z,t)$ is the actual ground-state coherence, $\Delta_{\mathrm{{HFS}}}\approx2\pi\times6.8\ \mathrm{{GHz}}$
is the hyperfine splitting between levels $|g\rangle$ and $|h\rangle$
and $K_{z0}=\sqrt{{\omega_{0}^{2}/c^{2}-k_{x}^{2}-k_{y}^{2}}}-\omega_{C}/c$
($\omega_{C}$ - coupling field frequency, $k_{x},\ k_{y}$- transverse
spatial components of the signal beam with respect to the coupling
beam; for our case $k_{y}=0$ and $ck_{x}/\omega_{0}\approx8\ \mathrm{mrad}$).
Then, the light-coherence evolution is given by following coupled
equations written in the frame of reference co-moving with the optical
pulse:
\begin{equation}
\frac{\partial\check{\rho}_{hg}(z,t)}{\partial t}=\frac{i}{\hbar}\frac{\Omega^{*}(t)dA(z,t)}{4\Delta-2i\Gamma}-\frac{1}{2\tau}\check{\rho}_{hg}(z,t)+i\delta_{\mathrm{tot}}(z,t)\check{\rho}_{hg}(z,t),
\end{equation}
\begin{equation}
\frac{\partial A(z,t)}{\partial z}=-i\frac{\hbar\Omega(t)\check{\rho}_{hg}(z,t)/d+A(z,t)}{2\Delta+i\Gamma}\frac{\Gamma}{2}gn(z),
\end{equation}
where $1/(2\tau)=|\Omega(t)|^{2}\Gamma/(8\Delta^{2}+2\Gamma^{2})$
is decoherence caused by radiative broadening, $\delta_{\mathrm{tot}}=\delta_{0}+\delta_{\mathrm{acS}}+\delta_{\mathrm{SSM}}+\delta_{Z}$
is total two-photon detuning including ac-Stark shift caused by control
beam $\delta_{\mathrm{acS}}=|\Omega(t)|^{2}\Delta/(4\Delta^{2}+\Gamma^{2})$,
SSM and spatially varying Zeeman shift $\delta_{Z}=\mu_{0}g_F m B_{0}+\beta z$
, with $F=2$, $m=2$ and $g_{F=2}=1/2$ , caused by linearly varying external magnetic field $B=B_{0}+\frac{\beta}{\mu_{0}}z$, where $\mu_{0}$ is
the Bohr magneton. The atomic concentration is denoted by $n(z)$ and
we define the ensemble optical depth as $\mathrm{OD}=\intop gn(z)\mathrm{d}z$.
In practice the value of $\delta_{0}$ is chosen to cancel out the
light shift caused by the control field: $\delta_{0}=-\delta_{\mathrm{acS}}$.

\subsection*{Spectral resolution}

The spectral resolution of the device is limited by finite duration
$T$ of the measurement window combined with the exponential decay
of the atomic coherence caused by the control field. One could consider
that upper limit for $T$ is given by combination of the bandwidth
$\mathcal{B}$ and the control field chirp $\alpha$ by $T_{\mathrm{max}}=\mathcal{B}/\alpha$
as for $\alpha T>\mathcal{B}$ a monochromatic input field $\tilde{A}(\omega)=\delta(\omega)$
lies outside the inhomogeneously broadened absorption spectrum. However,
in the usual operation regime we set $\alpha\ll\mathcal{B}^{2}$ and
to maintain high initial efficiency we always have $\tau<T_{\mathrm{max}}$.
In this regime the finite atomic coherence lifetime $\tau$ limits
the available measurement time $T$ which we set to be $T=\tau$
to maintain high overall efficiency $\bar{\eta}$. To estimate the
resolution accounting for both $\tau$ and $T$ we calculate the power
spectrum of a monochromatic input pulse with exponentially decaying
amplitude $A(t)=(\Theta(t)-\Theta(t-\tau))\exp(-\frac{t}{2\tau})$:
\[
|\tilde{A}(\omega)|^{2}\propto\frac{1+e-2 \sqrt{e} \cos (\tau  \omega )}{4\tau^{2}\omega^{2}+1},
\]
and define the spectral resolution $\delta\omega$ as FWHM of the
power spectrum $|\tilde{A}(\omega)|^{2}$. We numerically find $\delta\omega/2\pi\approx0.78/\tau$.

\subsection*{Group-delay dispersion estimate}

By imposing the parabolic phase shift onto the atomic ensemble, we
imitate temporal imaging setups that use group-delay dispersion in
chirped fiber Bragg gratings (CFBG), or just fibers, to achieve large
group delays. The temporal propagation length we achieve in our setup
amounts to $f_{t}=9600\ \mathrm{{s}}$ which corresponds to a GDD
of $25\ \mu\mathrm{{s}}^{2}$ over our 1 MHz bandwidth. To achieve
such GDD, one would need $10^{12}\ \mathrm{{km}}$ of typical telecom
fiber (GDD $25\ \mathrm{{ps}^{2}/km}$) or billions of commercially
available CFBGs (GDD $\sim10^{4}\ \mathrm{{ps}^{2}}$).
\subsection*{Simplified protocol}
\begin{figure}
    \centering
    \includegraphics[width=\linewidth]{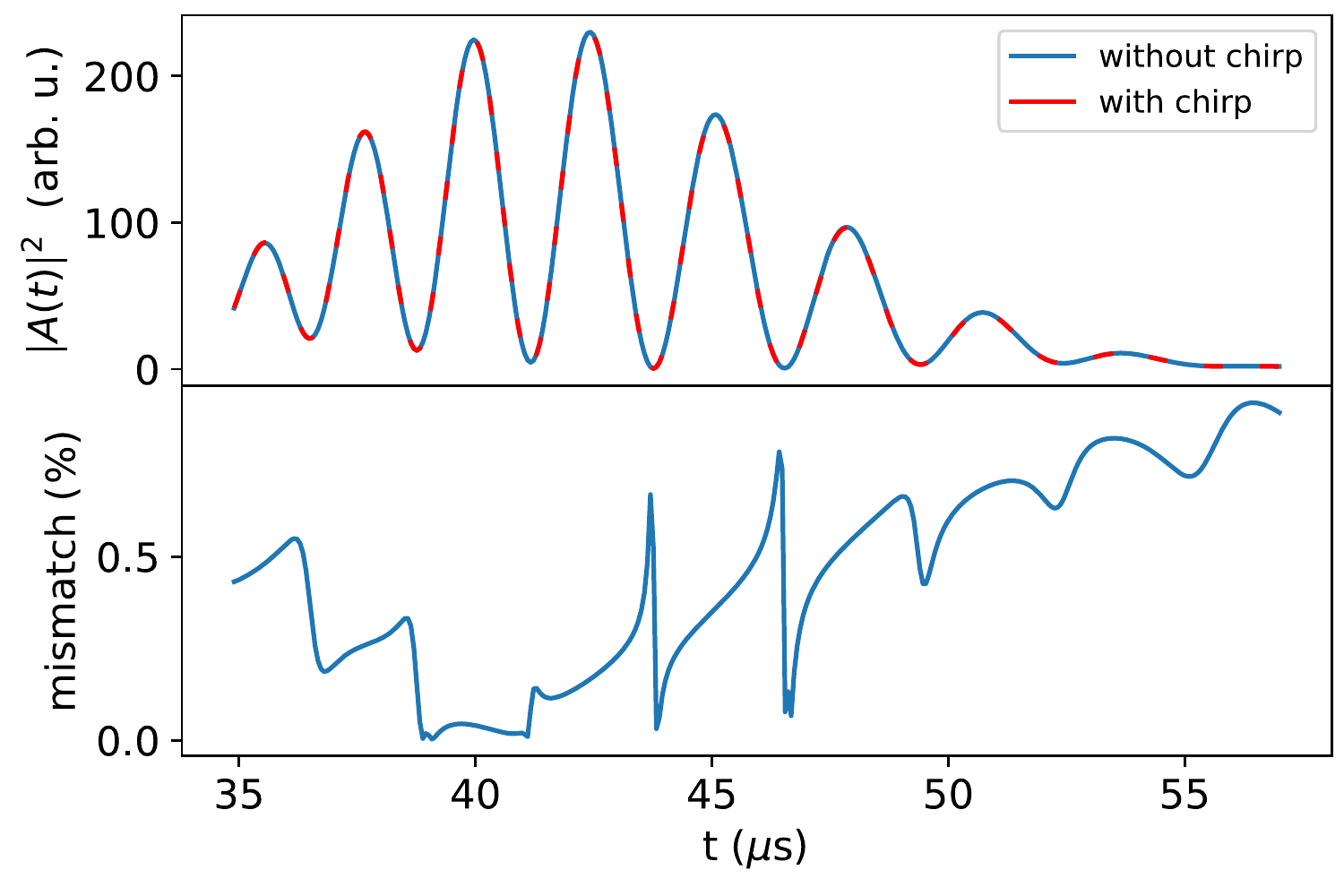}
    \caption{Simulated comparison of the output signal, as in Fig. 3(c) of the main Article with and without the chirp of the readout field. We observe no more than 1\% difference, varying that omitting the final chirp for the purpose of observing proper intensity profile at the output is well justified.}
    \label{fig:chirp_read_comp}
\end{figure}

In the main Article we mentioned that in practice the temporal far-field imaging sequence we do not implement the chirp at readout, as it would not change the final intensity. While this is true in the ideal case, here we also want to argument that in our particular implementation it bear almost no difference. In particular, then only reason the difference may arise is a relative change in the single-photon detuning $\Delta$. Here we simulate the protocol with and without the chirp and in Fig. \ref{fig:chirp_read_comp} we observe that the difference is minimal, staying always below 1\%.

\subsection*{Experimental setup details}

\subsubsection*{Magnetic field}
To determine the quantization axis the atomic cloud is kept in external constant $\sim1$ G magnetic field along the cloud, generated by external Helmholtz coils. The gradient of magnetic field for GEM is generated by a pair of rounded square-shaped coils in opposing configuration connected in series. The coils are made of 9 turns of copper wire wound over a base with a side length $d=10$ cm. The separation between the coils is $L=17$ cm. This configuration provides almost linear magnetic field gradient of value 0.08 G/A/cm over the 10 mm long cloud. The coils current and thus the gradient can be quickly switched to the opposite value using an electronic switch capable of connecting either constant current source (typ. 15A) in either direction or providing 160V for fast current reversal. This is accomplished using MOSFETs in H-bridge topology. The high switching speed (4.3 A/$\mu$s corresponding to $0.35$ G/cm/$\mu$s) equals 160 V divided by inductance of the coils. 

The instantaneous current in the coils is measured using \mbox{LEM-LA 100-P} current transducer with sub $\mu$s response time. The apparent overshoot after the current reversal is in fact a start of a combination of oscillation of LC circuit formed by the coils inductance and parasitic capacitance of the switch and an exponential decay of the current to a new steady state value. When the control field is not applied the value of the magnetic field gradient $\beta$ controls only the speed of shifting the atomic coherence in the $K_z$ space. Therefore, by applying the control after a short delay from the end of the switching state (when the overshoot is mostly present) we avoid the unwanted effect of readout efficiency change due to different instantaneous bandwidth $\mathcal{B}$. The residual variation of $\beta$ during the readout process slightly affects the temporal profile of readout speed, which manifests as a chirp of the expected intensity oscillations of the output signal, as in Fig. 3(d) of the Article. 

The magnetic gradient overshoot is in reality much smaller than estimated from the instantaneous current. We attribute this to eddy currents created in the metal parts of the magneto-optical trap apparatus occurring during switching of the coils which virtually compensate the amount of gradient change due to the overshoot.

\subsubsection*{Filtering system and noise characterization}
\begin{figure}[!t]
\includegraphics[width=1\columnwidth]{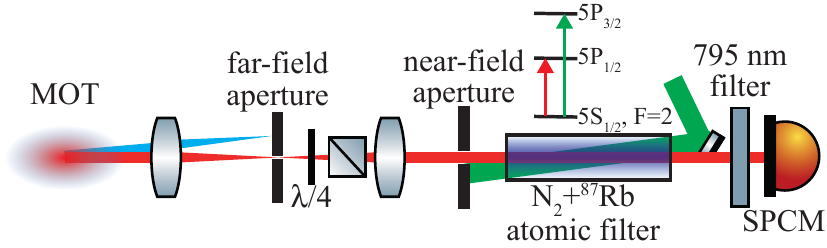}

\caption{Filtering setup. The signal separated from the coupling laser light
using a sequence of far field apertures, Wollaston polarizer, near
field aperture, optically pumped atomic filter and interference filter.
Transmission of the signal photons through this system amounts about
60\%.}
\label{fig:filtering}
\end{figure}
\begin{figure}[t]
\includegraphics[width=1\columnwidth]{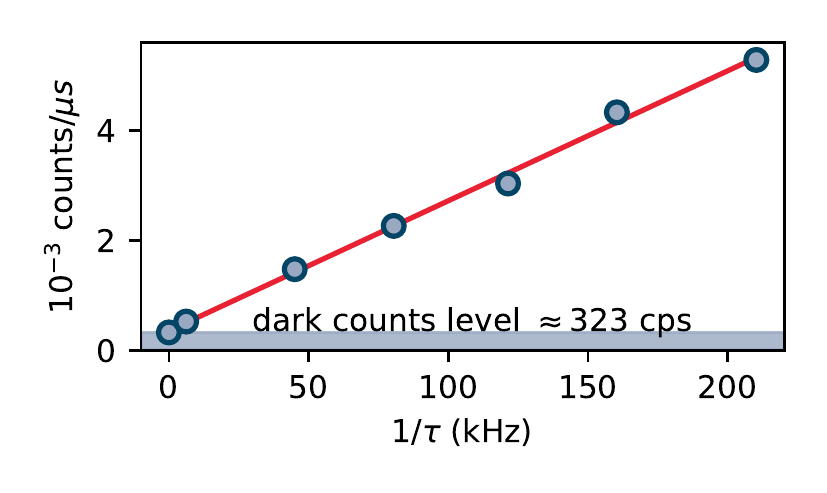}\caption{SPCM noise photons count rate versus decay rate of atomic coherence.
The slope of the fitted line amounts to 0.023 and can be interpreted
as average number of noise photons registered during readout process.}
\label{fig:noise}
\end{figure}
To minimize the noise we need to efficiently filter signal photons
from the control beam and other noise. For this purpose, we have built
a multi-stage filtering system (Fig. \ref{fig:filtering}). Firstly,
we filter most of control beam photons using far field aperture. Next,
we use the fact that they have orthogonal polarization to the signal
photons and we filter them using quarter-wave plate and Wollaston polarizer.
After that we use near field aperture to remove photons scattered
in other parts of MOT. Later, the glass cell containing Rubidium-87
pumped to $5S_{1/2},F=1$ state with 780 nm laser and buffer gas (nitrogen)
is used to filter out stray control beam light while preserving the
multimode nature of our device as compared to the cavity based filtering.
Finally we use a 795 nm interference filter to remove other frequency
photons, coming mainly from the filter pump. 

To estimate the amount of noise present in our experiment we use the same experimental sequence as in Fig. 3 but without any optical field at the input. We perform the experiment for different values of the control laser power $P$ and calculate the noise count rate for each point. 
In Figure \ref{fig:noise} we present measured noise count rate as a function of atomic coherence decay rate $1/\tau$, which is proportional to the power of the control beam $P$ (see Fig. 4(d) in the Article). The slope of the fitted line can be interpreted as average number
of noise photons registered during the typical $\tau$-long readout process. Note that
as we increase the coupling laser intensity, we register more noise
photons yet during a shorter window. This gives us a constant mean
photon number per readout. Thanks to our filtering system we achieved
the value of $\bar{{n}}_{\mathrm{{noise}}}=0.023$ which means, that
we register approximately 1 noise photon per 40 single experiments.
Simultaneously, the transmission of the signal photons amounts to
about 60\%, while the detection efficiency is $\sim$ 65\%.
With a typical memory process efficiency of 25\%, we obtain noise
per single photon sent to the device $\mu_{1}=0.23$, which corresponds
to $\mu_{1}=0.016$ per single mode (i.e. in a single temporal mode
storage experiment).The main limitation is still filtering of coupling
light, as witnessed by removing the atomic ensemble and still observing
the same noise level. That could be improved further by coupling the
signal to a single-mode fibre or using more efficient filtering.

\begin{figure}
\includegraphics[width=1\columnwidth]{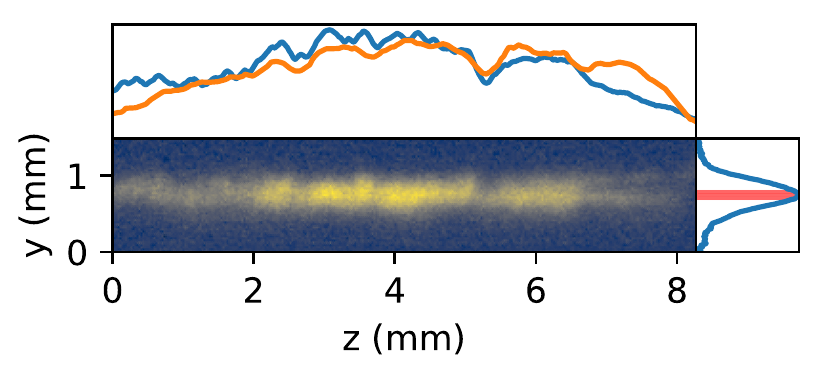}\caption{Fluorescence image of atomic cloud from the side. Center blue lines corresponds
to its integrals along $y$ and $z$ axis. Red area, to the right marks the part of atomic cloud illuminated with signal laser. The orange
line above presents concentration distribution along the $z$ axis corresponding
to the spectral profile in presence of magnetic field gradient.}
\label{fig:chmura}
\end{figure}

\begin{figure}[t]
\includegraphics[width=1\columnwidth]{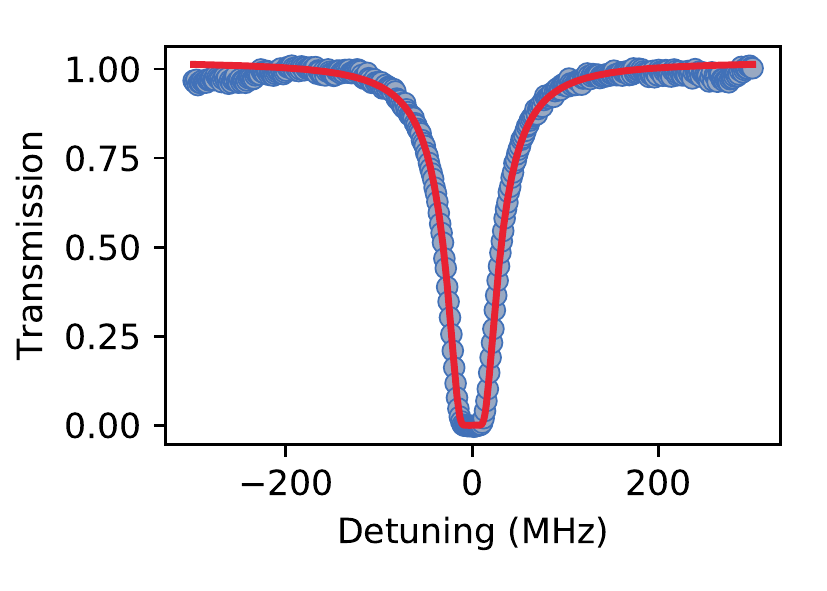}\caption{Single photon absorption profile of the signal laser. Fitted red
line corresponds to OD=76.}
\label{fig:od}
\end{figure}
\subsubsection*{Optical depth}
Figure \ref{fig:chmura} shows the image of atomic cloud from the side.
Atoms are formed into pencil shape area with diameter of about 0.5~mm. The signal laser diameter amounts to about 0.1 mm and illuminates
the middle of the atomic cloud, where the optical depth is the highest.
Figure \ref{fig:od} presents single photon absorption profile of the
signal. Fitting the saturated Lorentz profile, we estimated that optical depth
amounts to about 76.
\clearpage
\end{document}